\documentclass[twocolumn,superscriptaddress,floatfix,pra,aps]{revtex4}
\usepackage{graphicx,epstopdf,amsmath}

\begin{document}
\title{Vortex nucleation processes in rotating lattices of Bose-Einstein condensates ruled by  the on-site phases}
\author{D. M. Jezek}
\affiliation{CONICET - Universidad de Buenos Aires, Instituto de Física de Buenos
Aires (IFIBA). Buenos Aires, Argentina}
\author{P. Capuzzi}
\affiliation{Universidad de
  Buenos Aires, Facultad de Ciencias Exactas y Naturales, Departamento
  de F\'{i}sica, Buenos Aires, Argentina}
\affiliation{CONICET - Universidad de Buenos Aires, Instituto de Física de Buenos
Aires (IFIBA). Buenos Aires, Argentina}
\date{\today}
\begin{abstract}
  We study the nucleation and dynamics of vortices in rotating
  lattice potentials where weakly linked condensates are formed with
  each condensate exhibiting an almost axial symmetry.  Due to such a
  symmetry, the on-site phases acquire a linear dependence on the
  coordinates as a result of the rotation, which allows us to predict
  the position of vortices along the low density paths that separate
  the sites.  We first show that, for a system of atoms loaded in a
  four-site square lattice potential, subject to a constant rotation
  frequency, the analytical expression that we obtain for the
  positions of vortices of the stationary arrays accurately reproduces
  the full three-dimensional Gross-Pitaevskii results.  We then study
  the time-dependent vortex nucleation process when a linear ramp of
  the rotation frequency is applied to a lattice with sixteen sites.
  We develop a formula for the number of nucleated vortices which
  turns to have a linear dependence on the rotation frequency with a
  smaller slope than that of the standard estimate which is valid in
  absence of the lattice.  From time-dependent Gross-Pitaevskii
  simulations we further find that the on-site populations remain
  almost constant during the time evolution instead of spreading
  outwards, as expected from the action of the centrifugal force.
  Therefore, the time-dependent phase difference between neighboring
  sites acquires a running behavior typical of a self-trapping regime.
  We finally show that, in accordance with our predictions, this fast
  phase-difference evolution provokes a rapid vortex motion inside the
  lattice.  Our analytical expressions may be useful for describing
  other vortex processes in systems with the same on-site axial
  symmetry.

\end{abstract}
\maketitle

\section{ Introduction}

In the last decades the study of the dynamics of quantized
vortices in superfluid systems \cite{don91} has been an active
area of research.  In confined Bose-Einstein condensates (BECs)
\cite{dal99} the first experimental realization that succeeded in
generating and observing a vortex was performed in the year 1999
\cite{mat99}, and subsequently Anderson \textit{et al.} measured
the precession frequency of a vortex moving around the axis of a
harmonic trapping potential \cite{brian00}. Such experiments have
constituted the starting point of a blooming amount of works on
either static or dynamical configurations of vortices in BECs
within different types of confining potentials (see \cite{fet09}
and Refs. therein). Among them, the most commonly used trapping
potentials are the harmonic, quadratic plus quartic, and toroidal
ones,   to which   optical  lattices may  also be  superimposed.
The low density regions of such condensates, as for example the
central hollow of a torus, have shown to favor the pinning of
vortices. Whereas, for rotating systems with superimposed optical
lattices, the low density paths that connect the outside and
inside of the condensate, help to reduce the vortex nucleation
frequency.  The reason is that the energy barrier for a vortex to
enter the system becomes flattened when lowering such density.

The first experiments on producing arrays of many vortices were
performed with rotating harmonic traps  for large enough rotation
frequencies \cite{mad00,abo01,eng03} and rotating quadratic plus
quartic potentials for even larger frequencies
\cite{bret04,stock05}.  Then,  experiments dealing on quantized
circulation  have included toroidal trapping potentials
\cite{ryu07,eckel14,an21}.  In the more recent realization
\cite{an21}, a multiply quantized superfluid circulation has been
generated about the central density hollow  with a winding number
around the torus as high as $11$, which persisted for at least
$3$ seconds.  From the theoretical point of view, in such
toroidal trappings, stationary arrays of vortices have been
studied in rotating systems \cite{fet05,kim05} and in nonrotating
ones \cite{cap09,jez08}.  Finally, in rotating optical lattices,
some pioneer theoretical studies in searching equilibrium vortex
configurations have been carried out \cite{rei04,rei05,pu05}.
Such works dealt on vortices that became pinned on the density
minima, which was latterly experimentally observed by Tung
\textit{et al} \cite{tun06} in triangular and square lattices,
whereas for sufficient rotation intensity they reported a
structural crossover to a vortex lattice.  From that on, these
studies were extended to include other systems, e.g, mixtures of
species with more complex types of atomic interactions, as for
example the dipolar one \cite{kum19}. More recently, it has also
been analysed the consequences of including finite temperature
effects \cite{hass20}.

In an experimental work in a rotating lattice \cite{williams10},
the vortex nucleation process has been undertaken in a rather
distinct manner, since the optical lattice was subject to a
time-varying rotation frequency. In such work, it has been
observed that the number of nucleated vortices increased linearly
as a function of the frequency for barriers high enough so that
the system can be considered as formed by weakly linked
condensates (WLCs). In the present  work, although we use  a
simpler system, we work with WLCs and apply similar dynamical
conditions to those of the experiment, with the aim to better
understand such vortex nucleation process. An early attempt to
theoretically describe such results \cite{ka11} includes an
\textit{ad hoc} dissipation parameter to relax the system into
the equilibrium configuration, although in view of some
mismatches that appear  with respect to the experimental data, it
might not be the experimental case.

It is interesting to mention that density profiles similar to those of
lattice systems, can be attained in the supersolid phase of dipolar
BECs \cite{roccu20,gal20,an20} given that the formed particle droplets
are separated by low density valleys. Hence, in rotating condensates
the nucleation of vortices become more favored in the supersolid phase
than in the superfluid one.  In particular,  it was shown that low density
regions tend to reduce the energetic barrier for a vortex to enter the
system, thus lower  the nucleation frequency, and help in pinning
vortices in the interstitial zones between droplets \cite{gal20}.
We finally note that in an experimental work,  vortices have been recently 
observed in dipolar condensates \cite{klaus22}.

The goal of this work consists  in showing that for WLCs formed
in deep lattice potentials, the expressions for the
rotation-induced on-site phases  constitute a powerful  tool for
describing the nucleation and dynamics of vortices.  In Ref.
\cite{rot20} it has been observed that when a ring-shaped lattice
is subject to rotation, the induced on-site  phase of each
localized function (LF), and hence the associated velocity field,
depends on the geometry of the corresponding well. In some cases,
such  LFs phases acquire a simple expression, with a linear
dependence on the coordinates.  In this work we will obtain an
analytical formula for predicting the position of vortices on
rotating optical lattices by using such phases expressions.  The
formalism applies to square lattices for which the system can be
considered to be formed by WLCs with each of the condensates
having  an approximately axial symmetry around a direction
parallel to the rotation axis.  In particular, we will  focus our
study on four- and sixteen-site square lattices confined by a
harmonic potential.  For the four-site system the LFs are
obtained by a basis transformation from four Gross-Pitaevskii
(GP)  stationary states.  Considering  that the order parameter
can be written as a linear combination of the on-site LFs with
their corresponding phases, the location of vortices is
determined by searching the zeros of such an  order parameter.
The vortex positions between two neighboring sites can be easily
obtained by only considering both associated  LFs.   In doing so
we obtain the stationary array of vortices in the case of a
lattice of four sites, which show to be in very good accordance
with the vortex positions obtained by GP  simulations. In a
second step, we generalize such a  formula for a lattice which
does not exhibit a ring-type form, and hence the LFs cannot be
obtained  through a basis transformation.  Nevertheless, in order
to find the vortex positions, due to the sites symmetry, we can
assume that  the  LFs phases acquire the same analytic
expressions. 
Finally, we analyse the vortex nucleation process when  the
rotation frequency  varies linearly in time  from zero to the
radial trap frequency, reproducing the dynamical conditions
applied in the experimental device \cite{williams10}.  Although
we do not adjust  the size of the lattice and parameters to the
experimental ones, the present study allows us to analyse the
dynamical characteristics of a similar time-dependent nucleation
process.

The paper is organized as follows. In Section II, we describe the
theoretical framework for the four-site ring-shaped lattice and
analyse the spatial validity range of the phase expression for
the on-site localized function described  in Ref. \cite{rot20}.
In Section III, we obtain the stationary arrays of vortices for
the rotating ring-shaped lattice at two given frequencies.
Whereas, in Section IV, we study the vortex nucleation process
when applying  a time-linear ramp of the rotation frequency for a
sixteen sites square lattice. In particular, we determine the
number of nucleated vortices as a function of the rotation
frequency and analyse the involved vortex dynamics during such a
process. Finally, Section VI is devoted to the conclusions.

\section{ Theoretical framework}

\subsection{On-site localized functions for a ring-shaped lattice}

Given we consider high barriers between sites, a quite accurate
approximation of the order parameter is obtained by a superposition of
on-site LFs. In a previous work, it has been
described the method for obtaining such localized states  for a
rotating ring-shaped lattice system \cite{rot20}.  Summarizing, one
should first obtain the stationary states $\psi_n(\mathbf{r})$ by
solving the corresponding  GP  equation \cite{gros61},
\begin{equation}
\left[ \hat{H}_0 +
g \, N|\psi_n(\mathbf{r})|^2 - { \mathbf{\Omega}}\cdot {\hat{ L}} \right] \psi_n(\mathbf{r})=\mu _n \psi_n (\mathbf{r}),
\label{GProtstatic}
\end{equation} 
where $\hat{H}_0=-\frac{ \hbar^2}{2m}\nabla^2 + V_{\text{t}}$  with 
$V_{\text{t}}$ the trapping potential, $g= 4 \pi a \hbar^{2} / m $ is the 3D
coupling constant in terms of the s-wave scattering length $a$ of the
atoms, $\hat{L}$ is the angular momentum operator, and
$\mathbf{\Omega}=\Omega\hat{z}$ is the applied angular  rotation  frequency  around the $z$
axis.
For a nonrotating system, the index $n$ denotes the winding
number, where the velocity field circulation  is calculated along
a circle that passes by the links   around   the lattice.  Such a
winding number  is generated through a  phase imprinting  method
performed  on the order parameter before the minimization
procedure. For a number of sites  $N_s$, the possible
independent  states are restricted to values of   $n$  in the
interval $- ||(N_s-1)/2 || \leq n \leq ||N_s/2 ||$ \cite{cat11},
where $ ||.||$ denotes the standard  integer part. 
Whereas in a  rotating system, using the same phase imprinting
method, depending on the value of  $\Omega$,  the winding number
of  the outcome of the minimization can  change in $N_s$ units.
Hence, the different stationary order parameters can  still be
labeled  with  the same values of $n$,  restricted  to $-
||(N_s-1)/2 || \leq n \leq ||N_s/2 ||$ \cite{rot20}, although $n$
may not correspond to the actual winding number.  It has been
shown in Ref. \cite{cat11} that the stationary states with
different $n$ values are orthogonal.  Furthermore,  one can
define a set of $N_s$ orthonormal LFs, given by the following
basis transformation \cite{rot20}
\begin{equation}
w_k({\mathbf r})=\frac{1}{\sqrt{N_s}} \sum_{n} \psi_n({\mathbf r})
 \, e^{-i n\theta_k } \,,
\label{wannier} 
\end{equation}
where $\theta _k=2\pi k/N_s$. The index $k$ labels the site where
the function is localized.  For $N_s=4$,   the value  $k=0$
corresponds to the quadrant $x>0$ and $y>0$, and  $k$  increases
in the counterclockwise direction around the ring from $k=-1$ to
$k=2$.  It is important to note that the choice of the global
phase of each $\psi_n (\mathbf{r})$ can affect the localization
of the LFs.  A discussion of how to choose such  a phase in order
to achieve maximum localization is given in \cite{mauro4p}.  In
this work we set  $ \arg{(\psi_n (\mathbf{r}))}=0$  at the
bisector of the $k=0$ site. We note that, in contrast to the
nonrotating case, the on-site LF obtained through Eq.\
(\ref{wannier}), cannot be reduced to a real function.  Due to
the rotation, the wavefunctions $\psi_n(\mathbf{r}) $ have a
nonvanishing  velocity field within each site and hence  carry a
spatially inhomogeneous phase profile. This inhomogeneity in the
on-site phase is then transferred to the LFs through Eq.
(\ref{wannier}).

Finally, the order parameter can be approximated employing the LFs
 as
\begin{equation}
\psi ({\mathbf r},t) = \sum_{k} \, b_k(t) \, w_k ({\mathbf r})
 \,,
\label{orderparameter}
\end{equation}
with $b_k(t)=\sqrt{n_k(t)}e^{i\phi _k(t)}$, where $n_k= N_k/N$
with $ N_k$   the occupation number at the site $k$. We note that
the global time-dependent phase $\phi _k(t)$ does not represent
the actual phase in the $k$-site when $\Omega \neq 0$, but it
only takes into account its time dependence, while as we have
mentioned the spatial profile of the phase is carried out by the
complex function $w_k({\bf r} ) $.

\subsection{ The system }

The condensate is formed by Rubidium atoms  and the trapping potential  is given by,
\begin{equation}
  V_{\text{t}}({\bf r} ) = \frac{ 1 }{2 } m \left[
    \omega_{r}^2 r^2 
    + \omega_{z}^2 z^2 \right] 
  + 
  V_b \left[ \cos^2(\pi x/d)+ 
    \cos^2(\pi y/d)\right],
\label{eq:trap4}
\end{equation}
where $r^2=x^2+y^2$ and $m$ is the atom mass.  Hereafter, the
time, energy, and length will be given in units of
$\omega_r^{-1}$, $\hbar\omega_r$, and $l_r=\sqrt{\hbar/(m
\omega_r)} $, respectively. 

For the four site,  ring-shaped lattice we have chosen  the
following harmonic frequencies  $ \omega_{r}= 2 \pi \times 70 $
Hz and $ \omega_{z}= 2 \pi \times 90 $ Hz, and the intersite
distance  $ d = 3.9  l_r = 5.1 \mu$m.   The barrier height  of
the lattice  is fixed  at   $V_b = 25 \hbar\omega_r $   in order
to obtain a system of  four WLCs  for a  number of particles of
$ N=10^4 $  \cite{rot20}.

\subsection{ Validity range for the expression of the rotation induced
  on-site phases }

When the  ring-shaped lattice  is subject to rotation, the LFs
phases acquire a linear dependence on  the $x$ and $y$
coordinates, related to a homogeneous velocity field
\cite{rot20}. Such a velocity field profile is a consequence of
the almost axial symmetry of each localized on-site density
around an axis parallel to the $z$-direction. In Ref.
\cite{rot20}, an analytical expression for the phase on each site
has been obtained. In this section, we will show that such
expression is  valid  in a wide region that includes  the
straight segments that separates the  sites.

In particular, it has been shown \cite{rot20} that  the LF  for
the $k$-site can be written as \begin{equation}
  w_k(\mathbf{r},\Omega) = |w_k(\mathbf{r},\Omega)| e^{i
  \frac{m}{\hbar}
  (\mathbf{r}-\mathbf{r}_{\text{cm}}^k)\cdot(\mathbf{\Omega}\times\mathbf{r}_{\text{cm}}^k)},
\label{wlphase1} \end{equation}
where $\mathbf{r}_{\text{cm}}^k $ is the center of mass of the
localized density $|w_k( \mathbf{r}, \Omega )|^2$. Hereafter, for
simplicity, we will  omit writing   the implicit   LF dependence
on  $\Omega$.

Such a particular phase dependence on the coordinates has
important consequences on the vortex nucleation phenomenon, which
takes place along low density paths that connect  the lattice
junctions. Due to the discrete symmetry, we will concentrate
ourselves on a specific junction, but the results remain valid
for the other ones. In particular, we consider the junction that
separates the $k=0$ and $k=1$ sites, which lies along the
semiaxis $y>0$.  We first rewrite Eq. (\ref{wlphase1}) in terms
of the center-of-mass coordinates: $\mathbf{r}_{\text{cm}}^k =
(x_k , y_k , 0)$ as,
\begin{equation}
 w_k(\mathbf{r}) = |w_k(\mathbf{r})| e^{i \frac{m}{\hbar}(y x_k - x y_k) \Omega}.
 \label{wlphase2}
\end{equation}
For the trapping potential here considered, the coordinates of the
center of mass of the localized densities verify: $ x_1 = - x_0$ and
$ y_1 = y_0 $, and their absolute values may be taken equal to
$ d/2 $. Then, the neighboring LFs read,
\begin{eqnarray}
	w_0({ x, y, z }) &=& |w_0(\mathbf{r})| 
 \, e^{i A(-x + y) } \,,
\label{wannier0} \\
	w_1({ x, y, z  }) &=& |w_1(\mathbf{r})| 
 \, e^{ -i A(x + y) } \, ,
\label{wannier1}
\end{eqnarray}
where $A =  d  m\Omega/(2\hbar)$.

\begin{figure}[!h]
  \begin{tabular}{cc}
    \includegraphics[width=0.5\columnwidth]{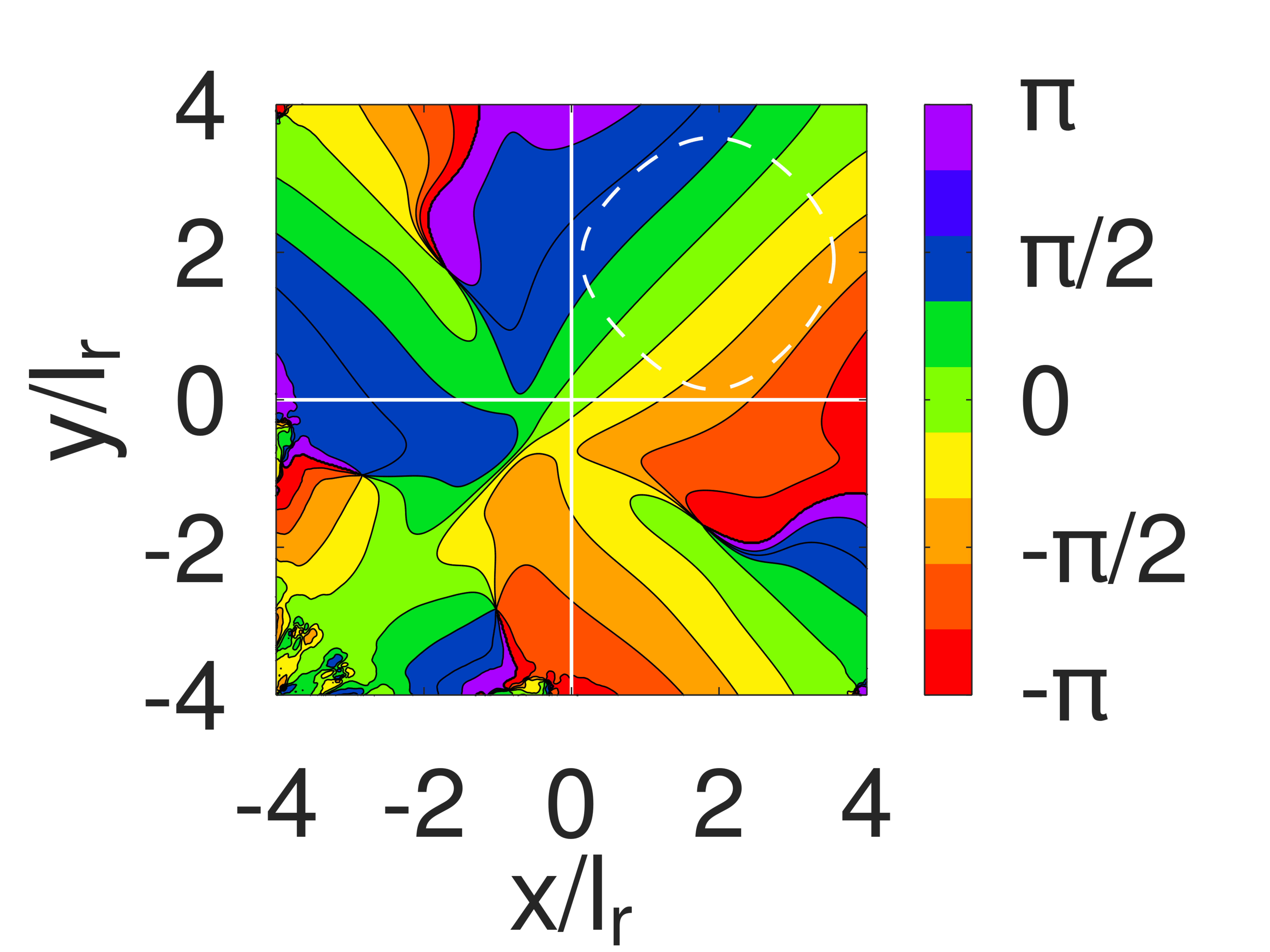}  &
                                                           \includegraphics[width=0.5\columnwidth]{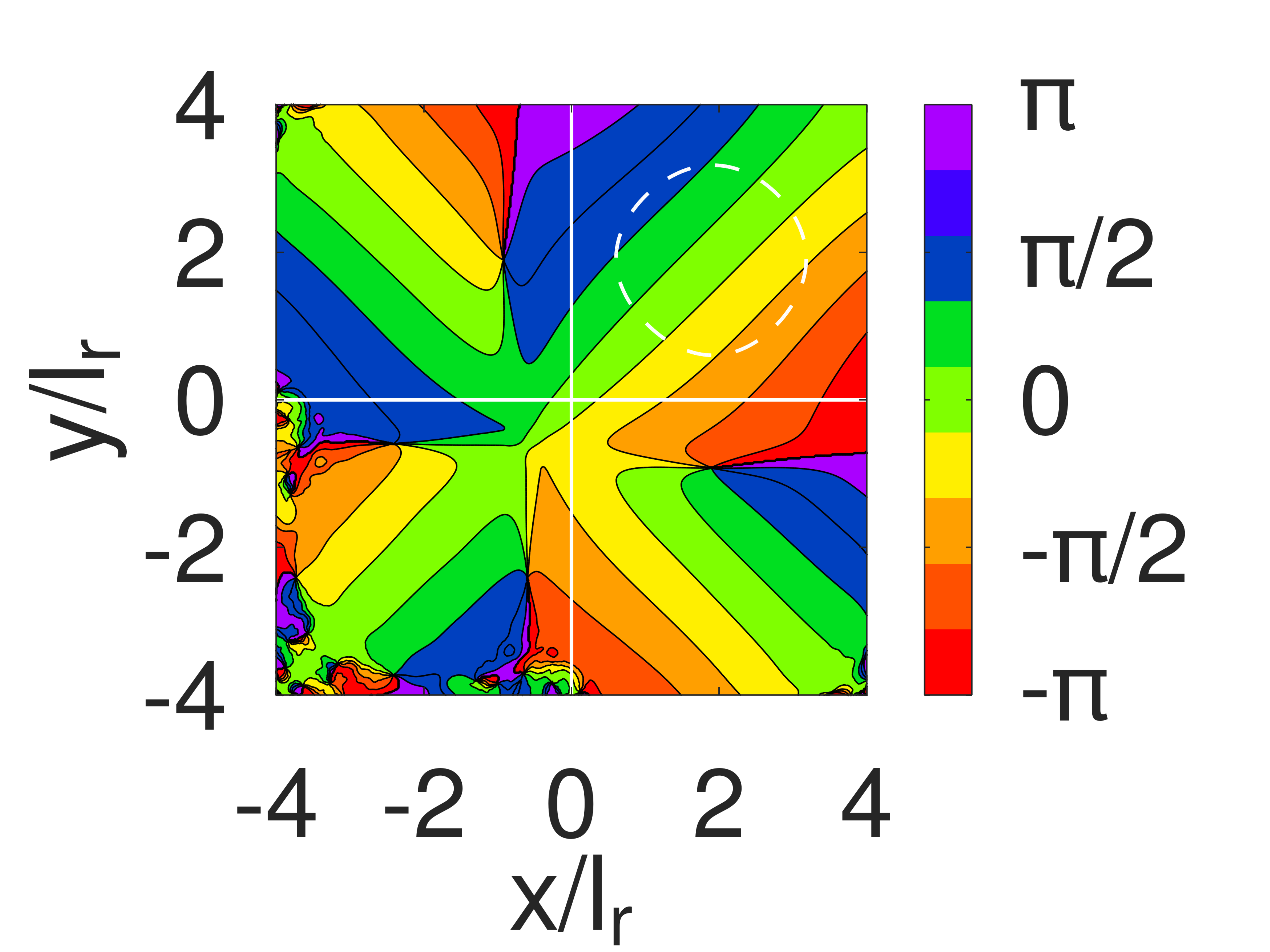}
    \\
 \includegraphics[width=0.5\columnwidth]{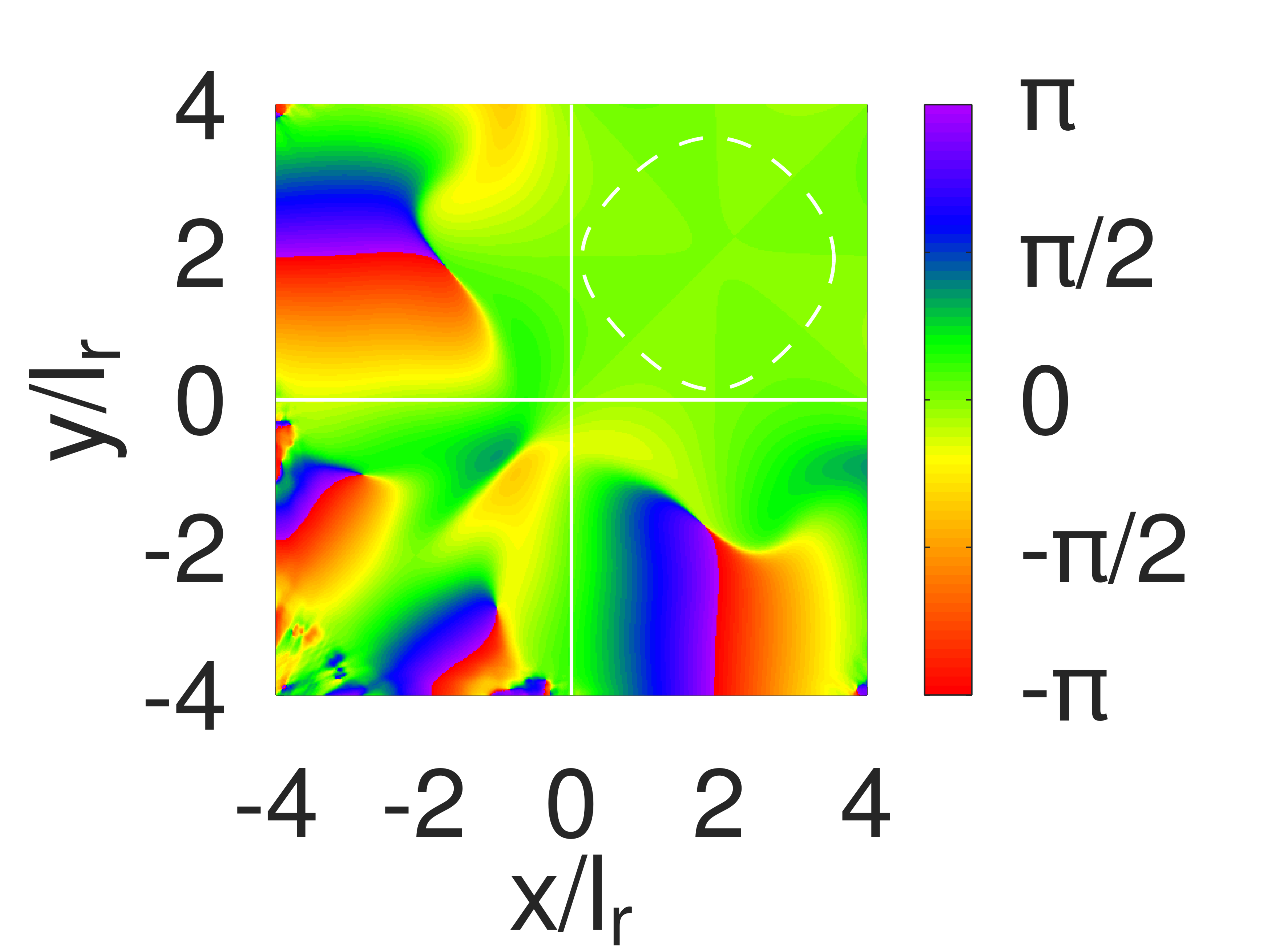} &
                                                       \includegraphics[width=0.5\columnwidth]{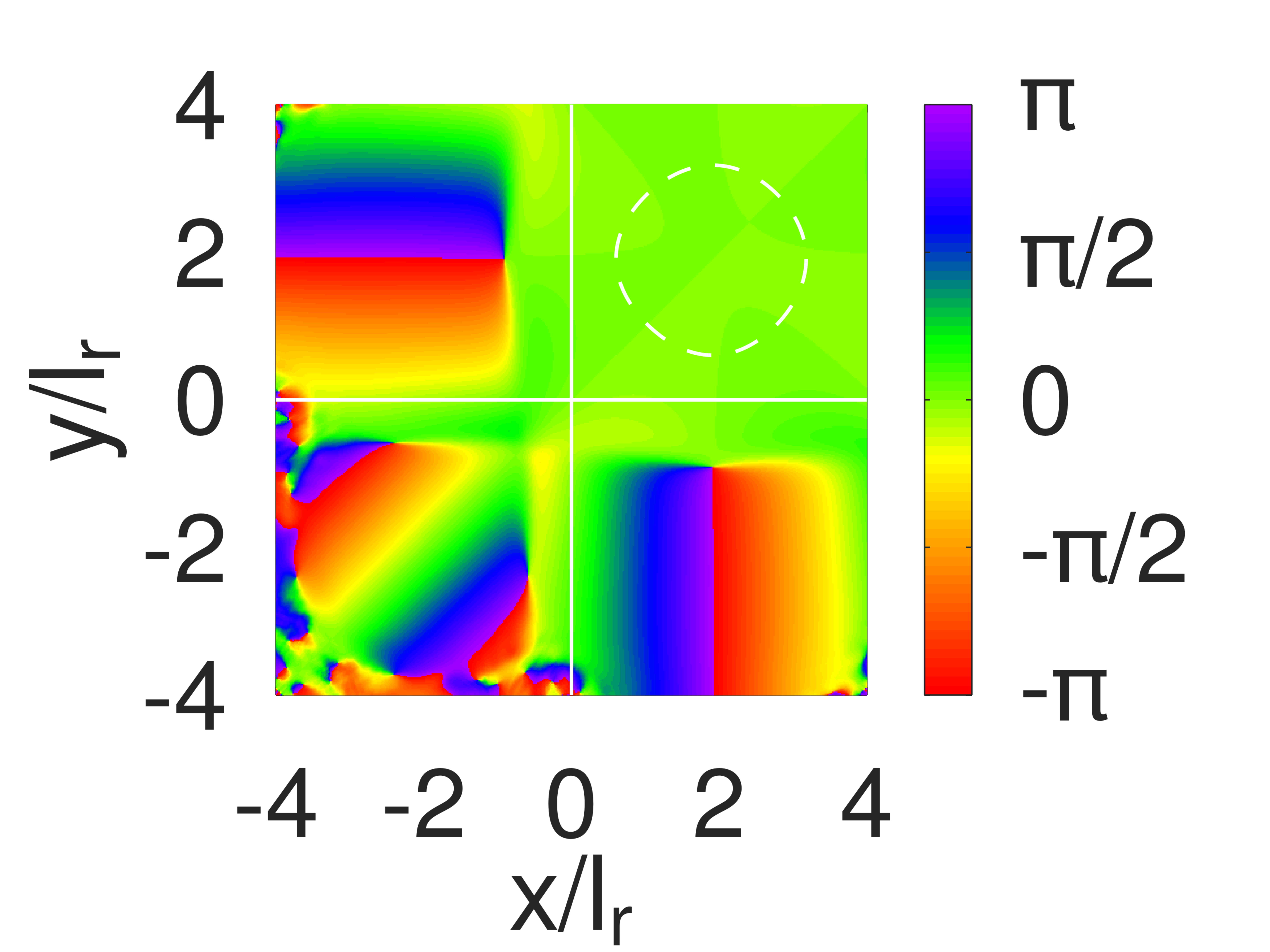}
 \end{tabular}

 \caption{\label{fig:phase} Top panels: Phase of $ w_0({ x, y, z_i })$
 from Eq. (\ref{wannier}) for $ \Omega/2 \pi = 28 $ Hz, at the
   planes $z_1=0$ and $z_2= 3 l_r $ in the left and right panels,
   respectively.  The form of the phase around the other sites is the
   necessary to ensure $ w_0({ x, y, z_i }) \perp w_k({ x, y, z_i })
   $, with $ k \ne 0$.   Bottom panels: Phase of $ w_0({ x, y, z_i }) \, e^{ -i
     A(-x + y) }$ for the same $ \Omega $ and $z$- planes of the top
   panels. The dashed lines indicate the isodensity
   contour of such LF with a value of $10^{-3}$ of the maximum
   density. The color scales correspond to $\arg{( w_0(\mathbf{r})) }$ and 
 $\arg{( w_0(\mathbf{r})\, e^{ -i
     A(-x + y) })}$ in the top and bottom panels, respectively.}
\end{figure}

 In the top panel of Fig. \ref{fig:phase} we show the LF phase profile
 for the $k=0$ site, given by Eq. (\ref{wannier}), at the $z=0$ and $z
 = 3 l_r$ planes, in the left and right panels, respectively.  In such
 panels it may be seen that the LF has a phase gradient with the shape
 of Eq. (\ref{wannier0}) within the site.  Moreover, in the bottom
 panels we depict such a phase minus its approximate analytic
 expression of Eq. (\ref{wannier0}). It may be seen that the resulting
 phase turns to be zero around the region determined by $x>0$ and
 $y>0$, for both $z$-planes.  We further note that given that the area
 of homogeneity is slightly larger than the mentioned quadrant, the
 actual phase $A(-x + y)$ remains valid surpassing the neighborhood of
 the junctions. Then, one can safely   use the order parameter of the
 type of Eq.  (\ref{orderparameter}),  with the expression for the LFs
 given by Eq.  (\ref{wlphase1}),  along the low density paths that
 separate the sites. As we will see, such an expression with the analytical phase  
 turns out to be crucial to correctly define the location of vortices.

\section{ Vortex nucleation on stationary states in a ring-shaped lattice }

\subsection{ Vortex nucleation on $ \psi_0(\mathbf{r})$ }

In order to analyse the appearance of vortices in the simplest
case, and in view that in current experiments the system is
initially in a vortex free state, we will first focus on the
$n=0$ stationary order parameter.  For a fixed $\Omega$ value
such a state, in terms of the LFs, with the on-site velocity
field provided by the rotation, is given by
\begin{equation}
\psi_0(\mathbf{r})=\frac{1}{\sqrt{N_s}} \sum_{k} w_k(\mathbf{r}) \,.
\label{ground}
\end{equation}
It is well known that a superfluid system is characterized by an
irrotational flow given by the condition on the velocity field
$\nabla \times \mathbf{v}(\mathbf{r})=0$. Since the vortex has a
nonvanishing velocity field circulation around its core, it
implies that the density should vanish at the coordinates of the
vortex line $ \mathbf{R}_v = (X_v,Y_v, z)$   to guarantee the
superfluid condition in every point of the fluid. Hence, we will
analytically obtain such  coordinates of the  vortex for the
junction around $x \simeq 0$ with $y>0$, by searching the points
that verify $\psi_0( \mathbf{R}_v$)=0, which  yields a  vanishing
density.

Using the analytic expression for the phases of the LFs and
retaining only the terms of the order parameter that include the
relevant localized states around that junction, we obtain the
vortex position by solving $\sqrt{N_s} \psi_0( \mathbf{R}_v)=
w_0(\mathbf{R}_v ) + w_1(\mathbf{R}_v )=0$ with the
approximations of Eqs. (\ref{wannier0}) and (\ref{wannier1}).
This  yields the following equation,
\begin{equation}
 |w_0( \mathbf{R}_v )| \, e^{i A(-X_v + Y_v) } + |w_1(\mathbf{R}_v )| \, e^{-i A(X_v+ Y_v) } = 0 \,,
\label{gr2}
\end{equation}
which  can be rewritten as,
\begin{equation}
 e^{-i A(X_v + Y_v) } \left( |w_0( \mathbf{R}_v )| \, e^{i 2 A Y_v } + |w_1( \mathbf{R}_v )|  \right) = 0 \,.
\label{gr3}
\end{equation}
Taking into account the real and imaginary part of the previous equation we further obtain,
\begin{equation}
 |w_0( \mathbf{R}_v )| \cos(2A Y_v) + |w_1( \mathbf{R}_v )| = 0 \,
\label{regr3}
\end{equation}
and
\begin{equation}
 \sin(2A Y_v) = 0 \,,
\label{imgr3}
\end{equation}
respectively.

Equation (\ref{imgr3}) implies $ 2A Y_v= \pi k' $, where $k'$ is
a natural number, whereas Eq. (\ref{regr3}) restricts such a
value to an odd number $ k'=2l+1 $ given that it should verify $
\cos(\pi k)= - |w_1({X_v, Y_v, z})| / |w_0({ X_v, Y_v, z })| =
-1$. We note that the condition for the absolute values of the
localized states, taking into account the symmetry,  is fulfilled
for $X_v=0$.
Whereas from  the other  conditions, one obtains the  expression
for the $Y_v$-coordinate along the low density path  between the
$k=0$ and  the $k=1$ sites,
\begin{equation} Y_{v}(l)= (2l+1) \, \, \frac{ \pi \hbar}{  m  d
\Omega} \,, \label{vor0} \end{equation} where  $l \ge 0 $  labels
the sequence of vortices that enter the system  from the $ y>0$
border of the lattice.  Moreover,  given the four-fold symmetry
of the lattice,  each $l$ value defines the positions of four
vortex lines: $(0, Y_v(l), z)$, $( Y_v(l),0, z)$, $ (0, - Y_v(l),
z)$, and $ (- Y_v(l),0, z)$, in the whole system.

\begin{figure}[!h]
\centering\begin{tabular}{cc}
\includegraphics[width=0.6\columnwidth]{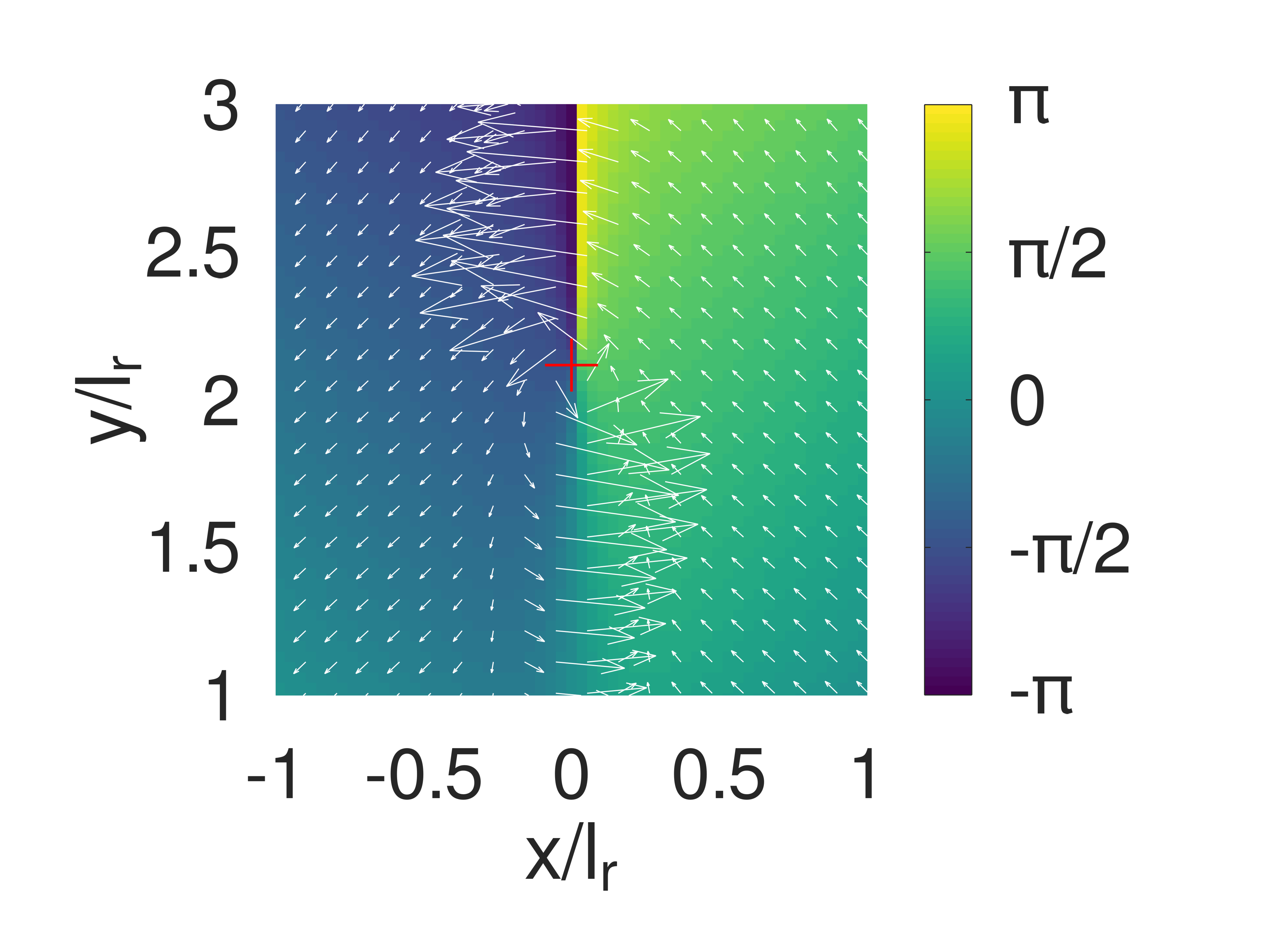}&
\includegraphics[width=0.35\columnwidth]{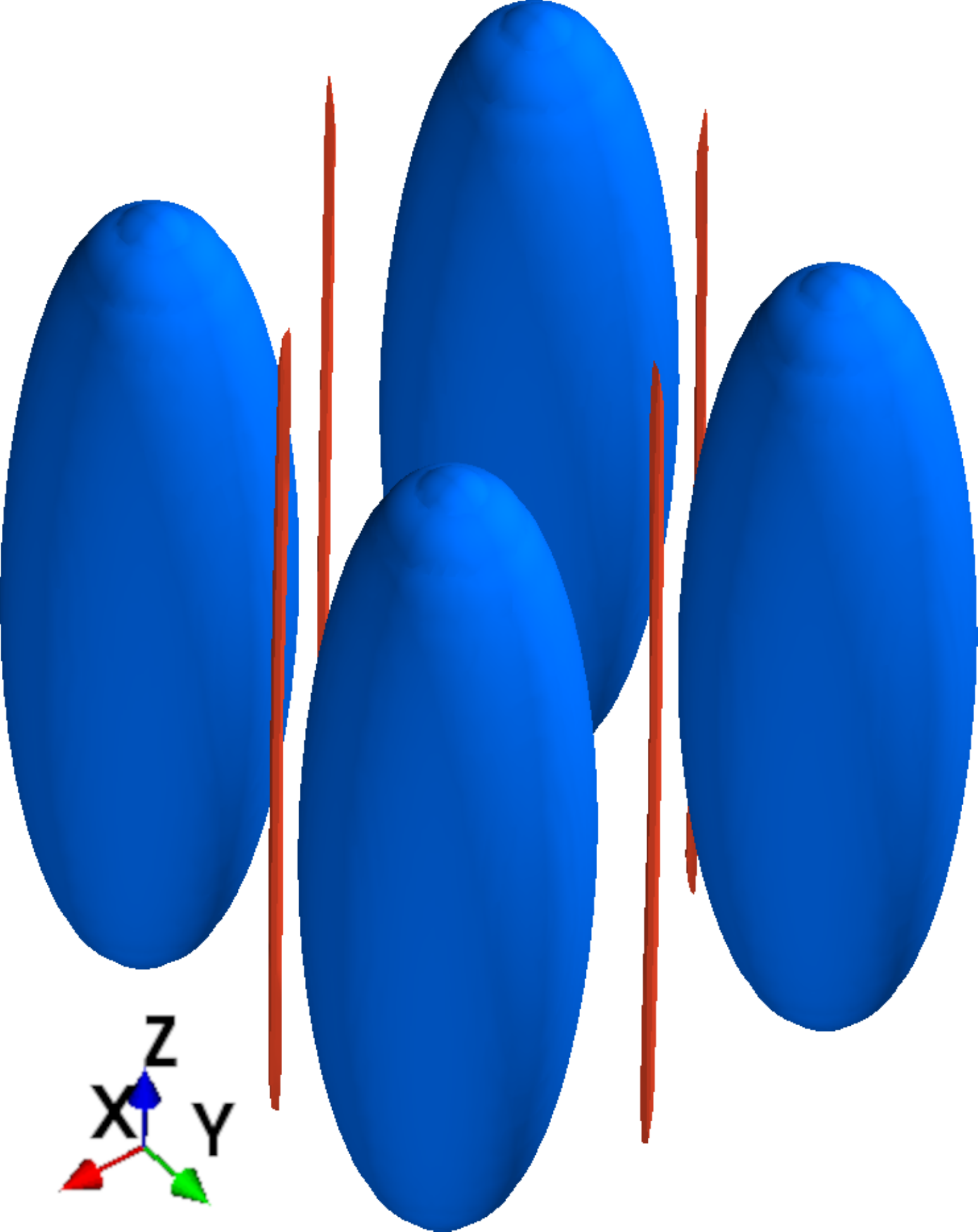}
\end{tabular}
\caption{\label{fig:velo0} Left panel:  Phase  (in colors) and
  velocity field (arrows) extracted from  the GP  order parameter
  $\psi_0( \mathbf{r} )$ around  the junction $y>0$ and at the
  $z=0$-plane, for $\Omega / (2 \pi) = 28 $  Hz. The  position of
  the vortex, obtained by the plaquette method, is marked with a
  red plus sign. Right panel: Three-dimensional configuration of
  the on-site condensates (blue) and the vortex lines (red)
  extracted  from GP simulations.  The color scale in the left
  panel corresponds to $\arg{( \psi_0(\mathbf{r})) }$. The border
  of the condensate in the right panel  is fixed at a density of
  $5\times10^{-2}$ of its maximum value.  }
\end{figure}

In the left panel of Fig.  \ref{fig:velo0} we show in colors the
phase and the corresponding velocity field  of  $
\psi_0(\mathbf{r})$ obtained from the GP  Eq.
(\ref{GProtstatic}), for  the  rotation frequency $\Omega / (2
\pi) = 28$ Hz.  There we mark with a red plus sign, a vortex
position,  which has been extracted from such a state  using  the
plaquette   method of  Ref. \cite{foster10}.  It  may be seen
that for  $l=0$, the estimate given by  Eq. (\ref{vor0}) yields $
Y_{v}(0)= 2.03 l_r $, which is in good accordance with the GP
result.  Such a position corresponds to a point of  the vortex
line $(0, Y_v(0), z)$.  In the right panel we show the full
system in three dimensions, with the four vortex lines.  We note
that such vortices coincide with straight lines parallel to the
$z$-axis as we have stated.

%

\subsection{ Vortex nucleation on the  stationary states }

We now extend the study to stationary states with arbitrary $n$
values. It is interesting to note that such states change their
relative energy values when varying the rotation frequency
\cite{rot20}, hence the ground state is not achieved at  a fixed
$n$ value for any $\Omega$.  By inverting the basis
transformation of Eq. (\ref{wannier}), such stationary states in
terms of the LFs acquire the form,
\begin{equation}
\psi_n( \mathbf{r} )=\frac{1}{\sqrt{N_s}} \sum_{k} w_k( \mathbf{r})
 \, e^{i n k 2\pi/N_s} \,,
\label{statn}
\end{equation}
for   $ n \in \{-1;0;1;2\} $.  In an analogous manner to the
described in the previous subsection, we can obtain the
$Y_v$-coordinate for the junction between $k=0$ and $k=1$ sites.
We then  search the solution of,
\begin{equation}
 |w_0(\mathbf{R}_v)| \, e^{i A(-X_v + Y_v) } + |w_1(\mathbf{R}_v)| \, e^{-i A(X_v+ Y_v) + i 2\pi {n}/{N_s} } = 0 \,,
\label{exc2}
\end{equation}
where $ \mathbf{R}_v = ({X_v,Y_v, z})$. From which  the following equation should hold,
\begin{equation}
 |w_0( \mathbf{R}_v)| \, e^{i (2A Y_v - 2\pi \frac{n}{N_s} ) } + |w_1
 ( \mathbf{R}_v)| \, = 0 \,,
\end{equation}
which is satisfied by $ 2 A Y_v - 2\pi n / N_s = \pi (2l+1) $.
Therefore, using that $N_s=4$, the $Y_v$-coordinate of the vortex is given
by,
\begin{equation}
Y_v(\Omega,n,l)= \left(\frac{n}{2}+ 2l +1\right) \, \frac{\hbar \pi}{ m d   \Omega} \,,
\label{vor1}
\end{equation}
where  $l \ge 0 $ labels the  different  sequences  of
vortices that enter the condensate, for the distinct $n$ values.

\begin{figure}
  \includegraphics[width=0.9\columnwidth]{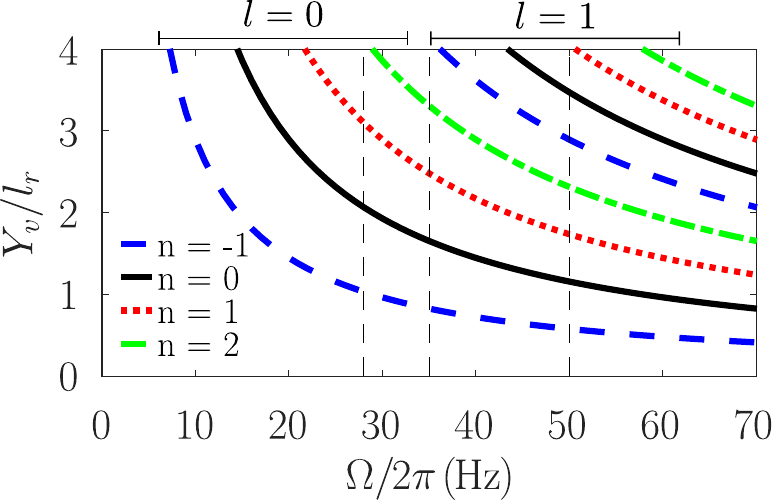}
  \caption{\label{fig:Yv} Vortex coordinate $Y_v$ given by Eq.
    (\ref{vor1}) as a function of rotation frequency $\Omega$. The
    different types of lines indicate the curves for the possible
     $n$ values, with $l=0$ and $l=1$ for the first and second
    sequence, respectively.    The vertical dashed lines indicate the 
 rotation frequency  values  $\Omega/2 \pi = \{28; 35; 50 \} $ Hz.}
\end{figure}
In Fig.  \ref{fig:Yv} we plot the vortex coordinate $Y_v$ as a
function of the rotation frequency, for all $n$ values and the
branches with $l=0$ and $l=1$. It may be seen that up to $\Omega/
2 \pi \simeq  35 $  Hz only one vortex is nucleated per junction,
whereas, for increasing  $\Omega$  a second vortex   per
junction, with $l=1$, enters the system for the stationary states
with  successive $n$ values.

\begin{figure}[!h]
 \tabcolsep=0pt
 \begin{tabular}{cc}
 \multicolumn{2}{c}{$\Omega=2\pi \times 28$Hz} \\
 \includegraphics[width=0.5\columnwidth,clip=true]{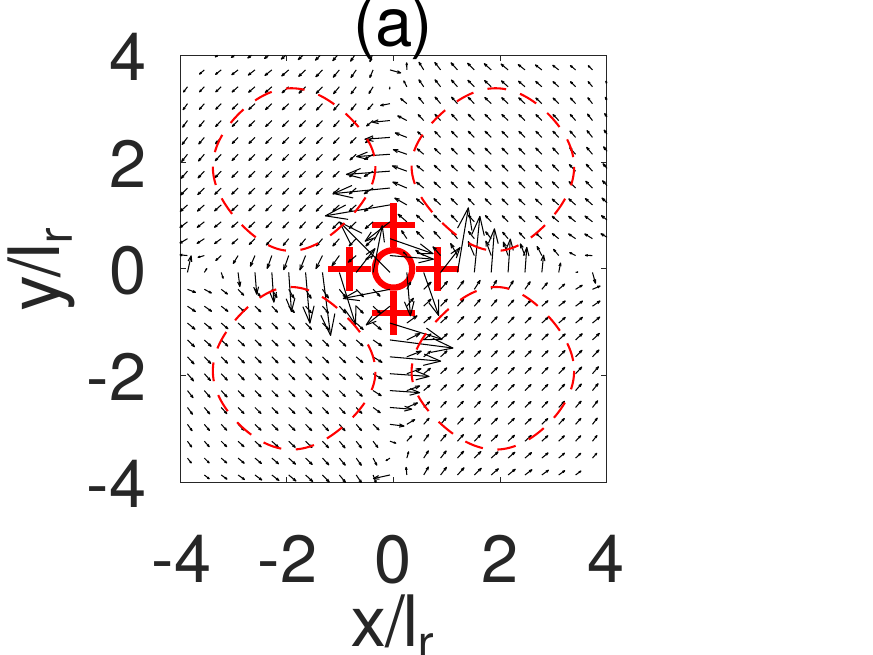}
 &
 \includegraphics[width=0.5\columnwidth]{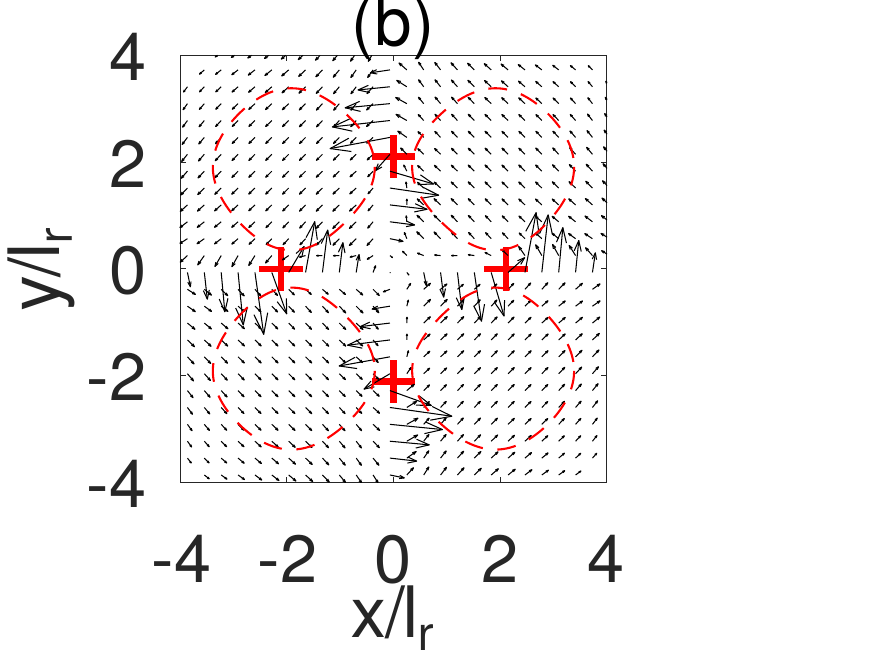} \\
 \\
 \includegraphics[width=0.5\columnwidth,clip=true]{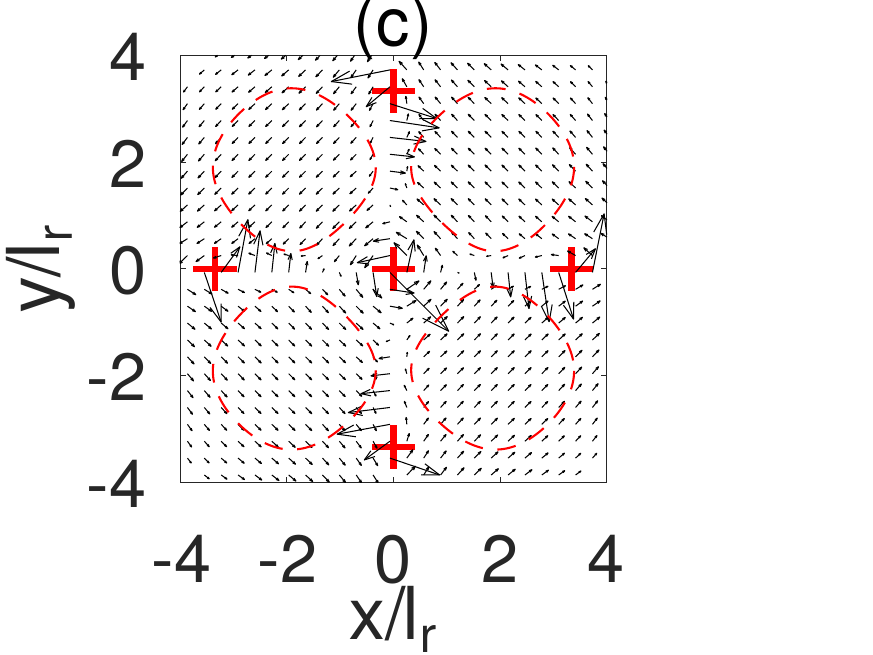} 
 &
 \includegraphics[width=0.5\columnwidth]{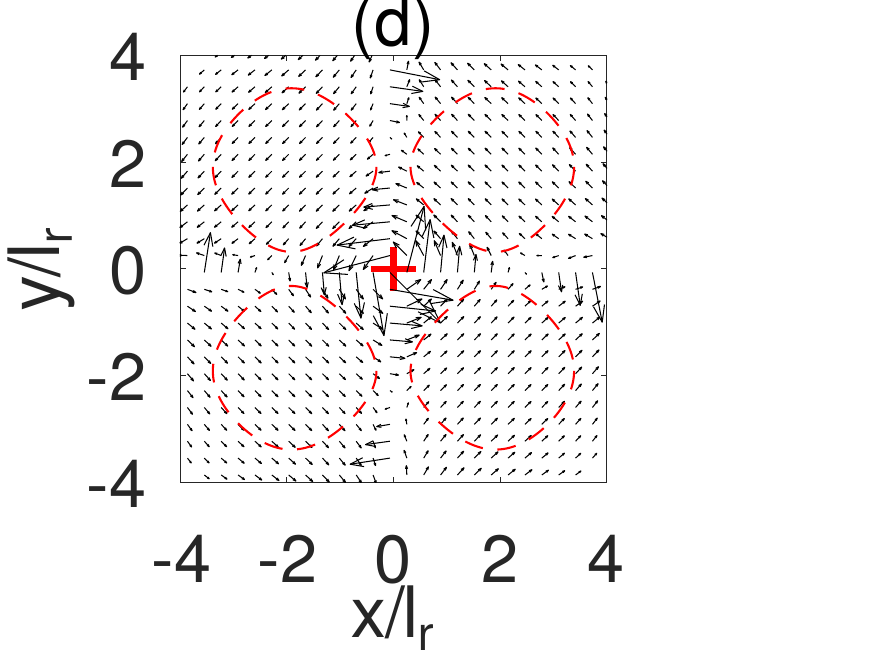}\\
 \multicolumn{2}{c}{$\Omega=2\pi \times 50$Hz} \\
 \includegraphics[width=0.5\columnwidth,clip=true]{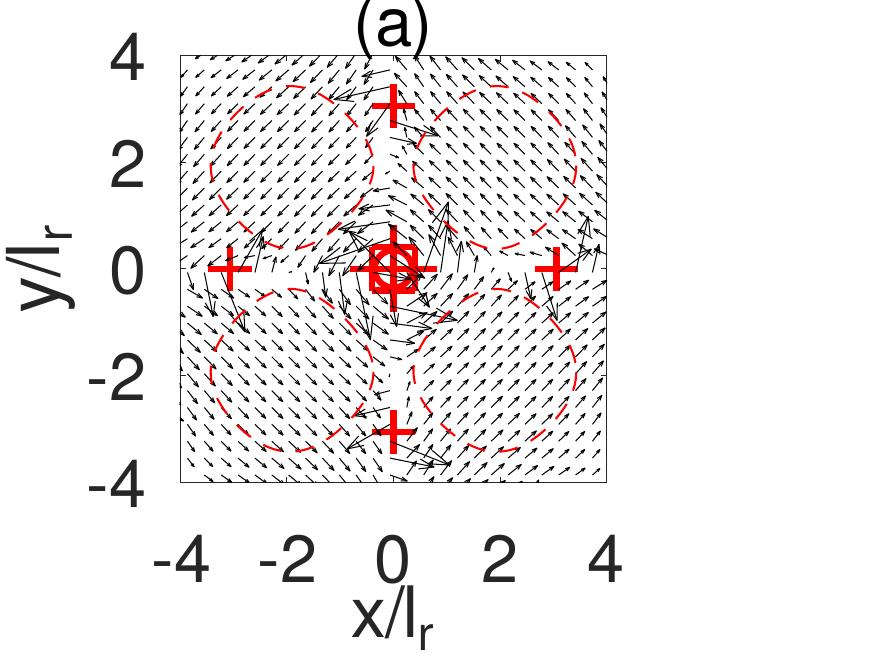}
 &
 \includegraphics[width=0.5\columnwidth]{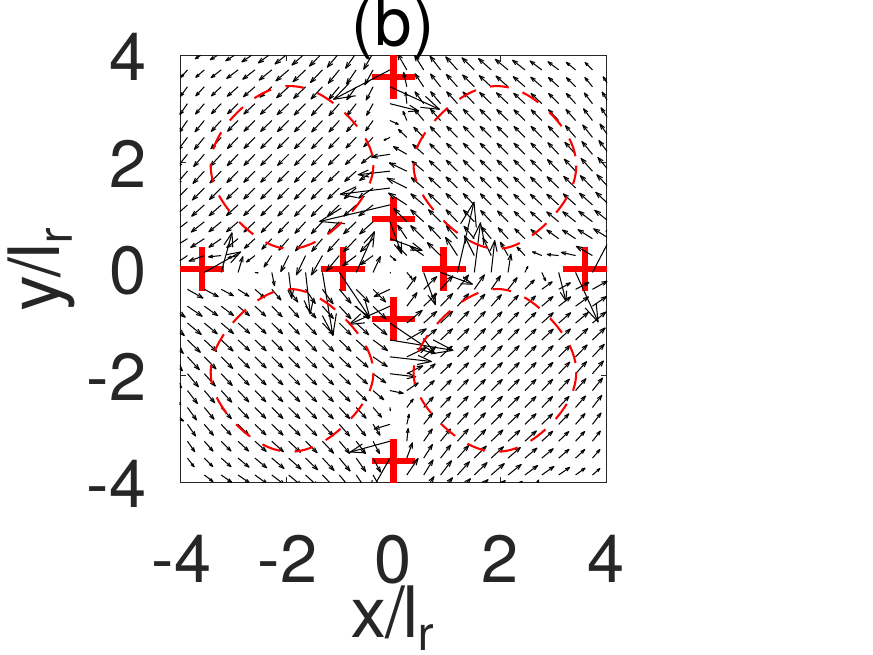} \\
 \\
 \includegraphics[width=0.5\columnwidth,clip=true]{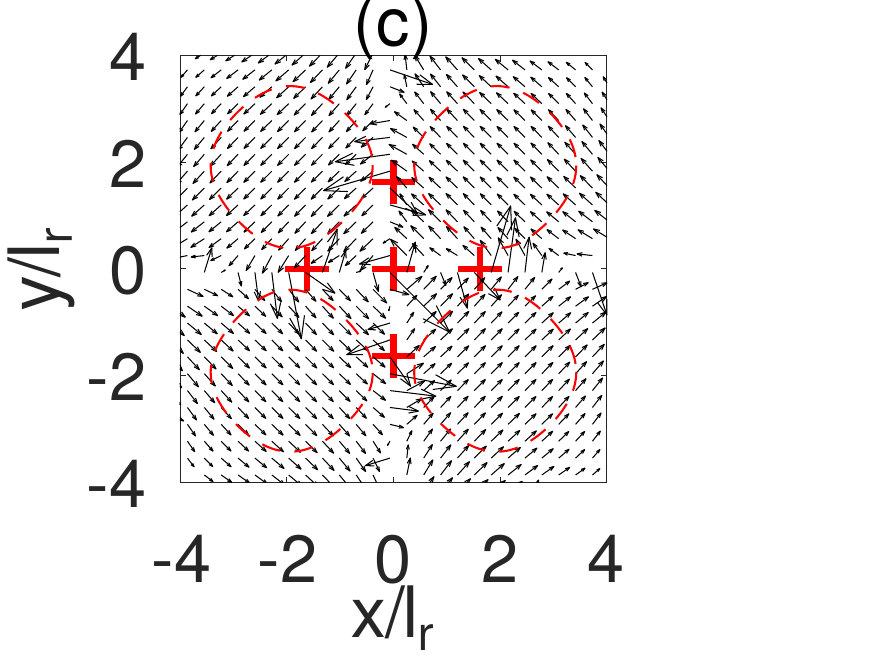} 
 &
 \includegraphics[width=0.5\columnwidth]{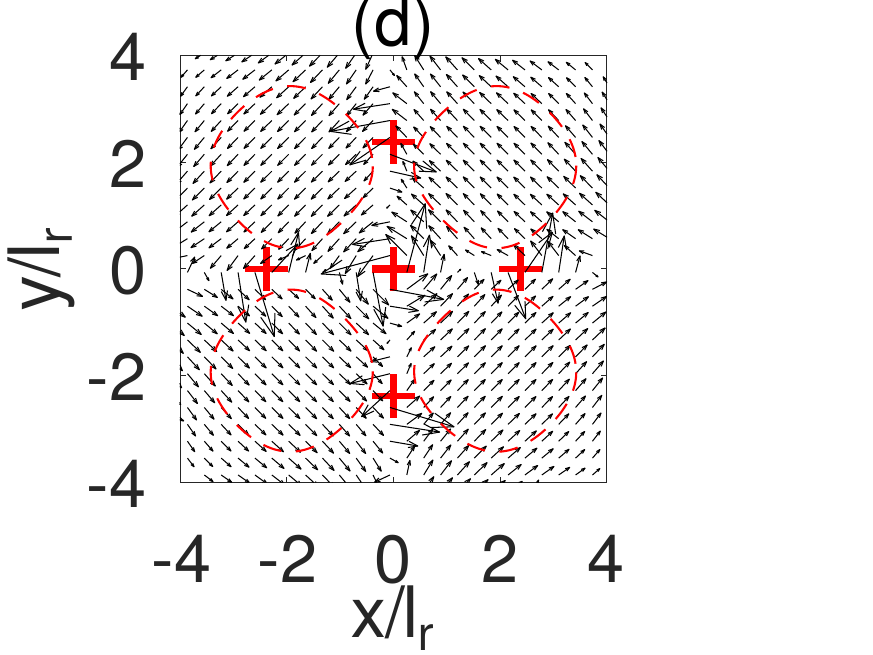}

 \end{tabular}
 \caption{\label{fig:velon} Stationary vortex arrays at the
   $z=0$-plane obtained from  $\psi_n(\mathbf{r} )$  of Eq. (\ref{GProtstatic}),  for  $\Omega / (2 \pi) = 28 $ Hz and
   $\Omega / (2 \pi) = 50 $ Hz.  The black arrows represent the
   velocity field  of the corresponding  order parameter.  Panels
   (a) to (d) of each frequency correspond to the different
   values  $ n = -1;0;1;2  $, respectively. The vortex  and
   antivortex  positions, obtained by the plaquette method, are
   marked with red  plus signs  and circles, respectively. Note
   that in the (d) panels, corresponding to $n=2$, the central
   plus signs denote  doubly quantized vortices centered at
   $x=y=0$.  The red dashed lines correspond to isodensity curves
   for a  value equal to $5\times 10^{-3}$ of the maximum
   density.   }
\end{figure}

In particular, in Fig. \ref{fig:velon} we show the arrays of
vortices in the $(x,y,0)$-plane for the different $ n  \in
\{-1;0;1;2\} $ values at two rotation frequencies, $\Omega/ 2 \pi
= 28$ Hz  and $50$  Hz.  The vortex positions of such arrays were
extracted from   $\psi_n (\mathbf{r})$  of  Eq.
(\ref{GProtstatic}), by means of  the plaquette method
\cite{foster10}.  The central vortices are related  to the
imprinted phases necessary  to obtain the different  $\psi_n
(\mathbf{r})$ states. The label $n=1$ ($n= -1$) corresponds to a
singly quantized  vortex (antivortex), $n=2$ to a doubly
quantized vortex, and $n=0$  has no central vortex. Such vortices
are present even for low frequencies.  In such a figure,  one can
observe that each, non central vortex position is in good
agreement with the expression of $Y_{v}$ which is depicted in
Fig. \ref{fig:Yv}, for the different $n$ and $l$ values.  For the
smaller  frequency the vortex positions are accurately reproduced
by the  $l=0$ sequence.  Whereas for the larger frequency, it may
be seen, that in  the panels (a) and (b) a second vortex per
junction  enters  the lattice  as expected  from  Eq.
(\ref{vor1}) for $l=1$.  By comparing the  vortex positions
obtained  from  both approaches one can infer that the estimate
derived  using the analytic expression for   the phases of the
LFs   has  shown  to be a very  reliable one.

It is interesting to recall  that  the  energy differences
between the  stationary  states $ \psi_n( \mathbf{r} )$  can
change their sign  for different frequencies.  In particular for
$ n = \{-1; 0; 1;2 \} $    we have obtained    $ E_n = \{-4.2; 0;
-6.7 ; -10.8 \}   10^{-4} \hbar \omega_r + 16.501048 \hbar
\omega_r $ and  $ E_n = \{-0.6; 0; 3.9; 3.3 \}  10^{-4}  \hbar
\omega_r +  15.213832 \hbar \omega_r $,   for  $\Omega/2 \pi = 28
$ Hz  and  $\Omega/2 \pi = 50 $ Hz, respectively. Hence, for  the
smaller  (larger) frequency  the ground state corresponds to
$n=2$ ($n=-1$).  By comparing  Eq.  (\ref{statn}) to  Eq.
(\ref{orderparameter}), for a given  frequency,  the stationary
states $ \psi_n( \mathbf{r} )$ have all the same population
numbers but differ in the  phase differences between neighboring
sites $\varphi_k=\phi_{k-1}-\phi_k= - n \pi/2$.  Then,  the
small energy differences, around $ 10^{-4}  \hbar \omega_r$,
between  the states  $ \psi_n( \mathbf{r} )$  can be  attributed
only  to  such  phase differences.

It is important to remark that the previous expression
(\ref{vor1})  for $Y_v$ does not describe the existence of the
central vortices since we have restricted the order parameter to
the superposition of only two neighboring LFs, while at the
origin, the four localized functions should be considered. We
thus have,
\begin{equation}
\psi_n(0,0,z)=\frac{1}{\sqrt{N_s}} \sum_{k} | w_k({ 0, 0, z})| \, e^{i n k 2\pi/N_s} ,  
\label{statn0}
\end{equation}
and since $ | w_k({ 0, 0, z})| = | w_0({ 0, 0, z})| $, one can
assure the presence of at least one vortex or antivortex along
the $z$-axis, for any $n \ne 0$, given that the previous
expression vanishes for such $n$  numbers.

\section{Time-dependent nucleation of vortices on extended square lattices }

We now consider a square lattice with a larger number of sites.
We are interested in systems with $N_s>4$ in order to analyse the
effects of the centrifugal force when the system is subject to a
time-linear ramp of the rotation frequency.  Such a number of
sites, which involves different absolute values of the
center-of-mass distance to the rotation axis, enables a possible
redistribution of particles during the time evolution.  We
further assume that each site has a high occupation number and
that the system forms a bosonic Josephson junction array of $N_s$
weakly linked condensates. In particular, we are interested in
studying the appearance of vortices when an initial vortex-free
ground state, for  $\Omega=0$,  is subject to the linear sweep
of  the frequency without losing the coherency between
neighboring sites during the evolution. Such a type of nucleation
process has been experimentally studied in Ref.
\cite{williams10}.

We first note that as the system is multiply connected, a single
vortex cannot be generated inside the lattice, since the winding
number along  any closed curve that connects  the WLCs through
the links cannot  be changed 
spontaneously.  As established in the celebrated Helmholtz-Kelvin
theorem \cite{lan95}, the winding number around a closed curve is
conserved during the time evolution if the superfluid condition is not
broken \cite{dam03}.  Therefore, the value of the velocity circulation
can only change when a vortex crosses such a  curve.  Hence, the vortex
nucleation process is originated by a vortex that penetrates the
system from the outside region, where the superfluid condition is
broken. It is worthwhile to mention that another possible process, not
observed in our case, could be the creation of a vortex-antivortex
pair within the lattice with a later departure of the antivortex.

In this study, we will consider each on-site condensate with the same
axial symmetry as has been assumed in the previous sections.  Although
obtaining complete approximate expressions for the on-site LFs in an
arbitrary square lattice does not constitute an easy task \cite{ka09},
we will see that in order to estimate the position of vortices only
the phase dependence on the coordinates is needed. Due to this axial
symmetry, one can safely assume that the rotation should induce a
linear phase with the same analytical expression as that of Eq.
(\ref{wlphase1}) on each LF.

On the other hand, taking into account the time-dependent character of
the present study, the LFs, the populations and on-site global phases could
evolve in time and then, the order parameter should be approximated
by,
\begin{equation}
\psi( \mathbf{r},t )=\frac{1}{\sqrt{N_s}} \sum_{k,p=1}^{ \sqrt{N_s}} w_{k,p} ( \mathbf{r},\Omega(t)  )
 \, e^{i \phi_{k,p} (t)} \sqrt{n_{k,p}(t)} \,,
\label{orderp}
\end{equation}
where here  the indices $k$ and $p$ label the sites in the $x$
and $y$ directions, respectively. We number the sites starting  from the 
corner determined by the smallest  and largest,  $x$ and $y$ coordinates, respectively.
Such  indices run
from $1$ to $ \sqrt{N_s}$. The population in each site is given by   $ N_{k,p} = n_{k,p} N$.
  In what follows, for simplicity, we omit writing
 the $\Omega$ dependence on time.

 In order to find the vortex position between the $k-1$- and $k$-sites, for a
given $p$, one can reduce the sum of  Eq.  (\ref{orderp}) to the two  nearest
neighboring LFs, and hence $\psi( \mathbf{r},t ) \sqrt{N_s} $ can be approximated by,
\begin{equation}
 w_{k-1}( \mathbf{r},\Omega )
 \, e^{i \phi_{k-1} (t)} \sqrt{n_{k-1}(t)} + w_k(\mathbf{r}, \Omega)
 \, e^{i \phi_k (t)} \sqrt{n_k(t)}  \, ,
\label{zeron}
\end{equation}
where for simplicity we have omitted the index $p$.
 Approximating the localized function by
\begin{equation}
 w_{k}(\mathbf{r},\Omega) = |w_{k}(\mathbf{r},\Omega )| e^{i \frac{m}{\hbar}
(\mathbf{r}-\mathbf{r}_{\text{cm}}^{k})\cdot(\mathbf{\Omega}\times\mathbf{r}_{\text{cm}}^{k})},
 \label{localg}
\end{equation}
 and taking into account that $k$
increases for increasing $x$ values we have,
$\mathbf{r}_{\text{cm}}^k = \mathbf{r}_{\text{cm}}^{k-1} + d \hat{x}$,
where $d$ is the intersite distance,  we may write the phase factor of  $  w_{k}(\mathbf{r},\Omega) $ from  Eq.  (\ref{localg}) as,
\begin{equation}
(\mathbf{r}-\mathbf{r}_{\text{cm}}^k)\cdot(\mathbf{\Omega}\times\mathbf{r}_{\text{cm}}^k) =
(\mathbf{r}-\mathbf{r}_{\text{cm}}^{k-1})\cdot(\mathbf{\Omega}
\times\mathbf{r}_{\text{cm}}^{k-1}) + y \Omega d.
 \label{phaseg}
\end{equation}
By introducing  (\ref{localg}) and (\ref{phaseg})  in Eq. (\ref{zeron}) one obtains  the following condition
for the vortex position  $\mathbf{R}^{(k)}_v =(X^{(k)}_{v}, Y^{(k)}_{v}, z )$,
\begin{multline}
 \sqrt{n_k(t)}|w_k(\mathbf{R}^{(k)}_v,\Omega )| \, e^{i (\frac{m
  \Omega d}{\hbar} Y_v^{(k)} - \varphi_k(t) ) } \\ + 
 \sqrt{n_{k-1}(t)} |w_{k-1}(\mathbf{R}^{(k)}_v, \Omega )| \, = 0 \,,
\label{exc3}
\end{multline}
where $\varphi_k(t)= \phi_{k-1}(t) -\phi_k(t) $ is the phase
difference between such neighboring sites, with  $ -\pi < \varphi_k(t) < \pi $.
 The superscript  $(k)$ in the vortex position denotes it belongs to the 
 path
  parallel  to the $y$-axis,  that  separates the sites labeled  with $k-1$ and $k$.
  Then, from the
 imaginary part of Eq. (\ref{exc3}), one can derive the
expression for the vortex coordinate,
\begin{equation}
Y^{(k)}_{v}(t)= \left( \frac{ \varphi_k(t)}{\pi} + 
2l_k + 1 \right) \, \, \frac{ \pi \hbar}{ m d \Omega} \,,
\label{vortimel}
\end{equation}
where  $l_k$ is an integer number that labels the vortices
located along such a   path  parallel  to the $y$-axis.  If one
wants to calculate   $ Y^{(k)}_{v}(t)$  along the whole low
density  straight line that crosses the lattice, one should
incorporate in  Eq.  (\ref{vortimel}), the phase differences  $
\varphi_k(t)$ for all  the $p$-values.  However, due to the
four-fold symmetry of the square lattice one can restrict the
study to $ Y^{(k)}_v >0$, which is given for $p \le  \sqrt{N_s}/2
$ and  $l_k\ge 0$. 

Finally,  we note that,  from the real part of Eq. (\ref{exc3})
one should obtain the transversal coordinate $X^{(k)}_{v}(t)$ of
the vortex along such a straight path,  if  having at  hand $
|w_k(\mathbf{R}^{(k)}_v,\Omega )|$, which is  not the present
case.  Although possible variations on the site populations,
$N_k$ and $N_{k-1}$, could lead to slight shifts in such
transversal coordinate, this effect turns out to be almost
imperceptible due to the tight on-site localization, as we have
observed from our numerical GP simulations. Then, we will
consider such $x$ values constant along the straight lines
parallel to the $y$-axis.

In the following subsection we will show the usefulness of the
previous formula for determining the number of nucleated
vortices.

\subsection{Number of  nucleated vortices}

In what follows, we will be focused on the number of vortices
that become nucleated on the lattice as a function of $\Omega$,
which turns to be an increasing function of time.  Assuming the
vortices  are generated outside the lattice, one can obtain such
a  number by counting how many vortices enter into the low
density paths by what we will call entrances.  Taking into
account the four-fold symmetry, we will only consider the  $
Y^{(k)}_v>0$ case.  Being the first vortex labeled by $l_k=0$,
the number of vortices, for a given  $\Omega$, that enter by such
an entrance, reads $N^{(k)}_v = l_k^{(L)}+1$, with   $l_k^{(L)}$
the largest $l_k$-value.  The value of  $l_k^{(L)}$   as a
function of   $\Omega$  can be obtained  through Eq.\
(\ref{vortimel}) with the condition that $ Y^{(k)}_v$ is located
at the entrance,  hence $Y^{(k)}_v = d \sqrt{N_s}/2 $, where
$\sqrt{N_s}/2$ is the number of sites in the $y$-direction
(labeled by $p$), up to the $x$-axis.  Then,  replacing
$N^{(k)}_v = l_k^{(L)}+1$, we finally  obtain,
\begin{equation}
 N^{(k)}_v(t) = \left\lVert \frac{ m d^2  \sqrt{N_s}  \Omega} { 4 \pi \hbar} - \frac{ \varphi_k(t)}{2 \pi} + \frac{1}{2} \right\rVert \,.
\label{niv}
\end{equation}

Given that the time evolution depends on the way the rotation
frequency is varied,  we can only obtain bounds of the previous
expression taking into account $ -\pi < \varphi_k(t) < \pi $,
which yield,
\begin{equation}
N^{(k)}_{-}  = \left\lVert \frac{ m d^2  \sqrt{N_s} \Omega} { 4 \pi \hbar} \right\rVert \,,
\label{nivminus}
\end{equation}
and, 
\begin{equation}
N^{(k)}_{+} = \left\lVert \frac{ m d^2 \sqrt{N_s}  \Omega} { 4 \pi \hbar} \right\rVert + 1 \,,
\label{nivplus}
\end{equation}
for 
the lower and upper bounds, respectively.

Multiplying by the number of entrances around the lattice $ 4 (
\sqrt{N_s} -1) $, and approximating the integer part function by
a straight line that passes by the middle points of  the  steps,
we obtain  $ ||X|| \simeq X - 1/2 $, for $X>0 $,  we conclude
that the total number of vortices $N_v$ inside the lattice lies
between the following values,
\begin{equation}
 N^{(\pm)}_v = 4(\sqrt{N_s}-1)\left(\frac{ m d^2 \sqrt{N_s} \Omega} {4 \pi \hbar}\right) \pm 2( \sqrt{N_s}-1) \,,
\label{nv}
\end{equation}
which can be rewritten as,
\begin{equation}
N^{(\pm)}_v = \frac{ m d^2 (N_s - \sqrt{N_s} ) \Omega}{\hbar \pi} \pm 2 (\sqrt{N_s} - 1) \,.
\label{deltavf}
\end{equation}
Finally,  an estimate of such a  number of nucleated vortices can be obtained as the average
of
$N^{(\pm)}_v $ which yields,
\begin{equation}
N_v = \frac{ m d^2 (N_s - \sqrt{N_s} ) \Omega}{\hbar \pi} \, .
\label{nva}
\end{equation}

Consequently, the total number of vortices nucleated in the
lattice is a linear function of $\Omega$ in agreement with the
experimental observation of \cite{williams10}. We note that
Eq.~(\ref{nva}) differs from the best  estimate used in
\cite{williams10}, where it is assumed that the density of
vortices for a given $\Omega$, whether a lattice is present or
not, is $ n_v = m \Omega/\pi \hbar$. Then the   total number of
vortices follows the Feynman's   rule   \cite{fey55}  $N_v^{F} =
m \Omega R^2 / \hbar $, where $R$ is the Thomas-Fermi radius.  In
Ref.  \cite{williams10} the authors  claim  that, when working
with WLCs,  a good estimate for  their results  is given by a
straight line which is obtained by  using a fixed $
R(\Omega)=R(\Omega=0)$, instead of considering  an increasing
$R(\Omega)$.  Such an approximation  in terms of the number of
sites $N_s$ and the intersite distance $d$  may be rewritten as
$N_v^{F} = m \Omega d^2 N_s /   \pi \hbar $, given that  the area
$ \pi R^2$ should be replaced  by $N_s d^2$.  Here, by assuming
that the vortices are coming from outside the lattice, we have
obtained a smaller slope due to the subtraction of the term
$\sqrt{N_s}$ inside the parenthesis   of Eq. (\ref{nva}).

\begin{figure}[!h]
\includegraphics[width=\columnwidth]{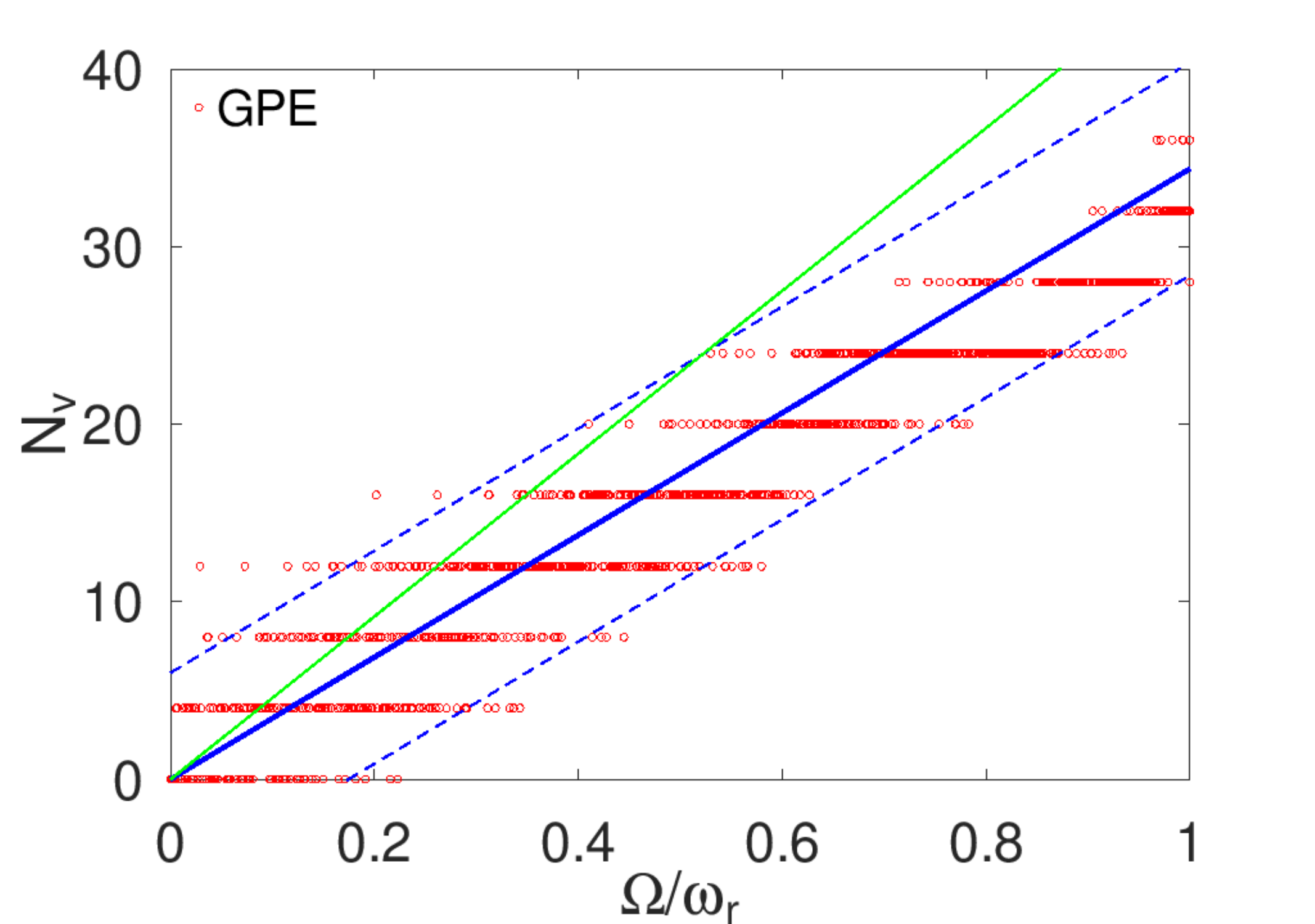} 
\caption{\label{fig:Vb75}Number of vortices $N_v$ as a function of the
  rotation frequency $\Omega$ for $V_b=75 \hbar\omega_r$. The red
  points correspond to time-dependent GP simulations, obtained  by solving Eq. (\ref{GProtdin}).
 Whereas the blue lines indicate our
  predictions, in particular,  the solid line indicates
  the mean value given by  Eq. (\ref{nva}) and 
 the dashed lines  correspond to the upper and
  lower bounds from Eq. (\ref{deltavf}). The green solid line, with the higher slope,
  corresponds to the estimate  $N_v^F = m \Omega d^2 N_s /   \pi \hbar $.  }
\end{figure}

\begin{figure}[!h]
 \centering
 (a)\\
 \includegraphics[width=0.85\columnwidth]{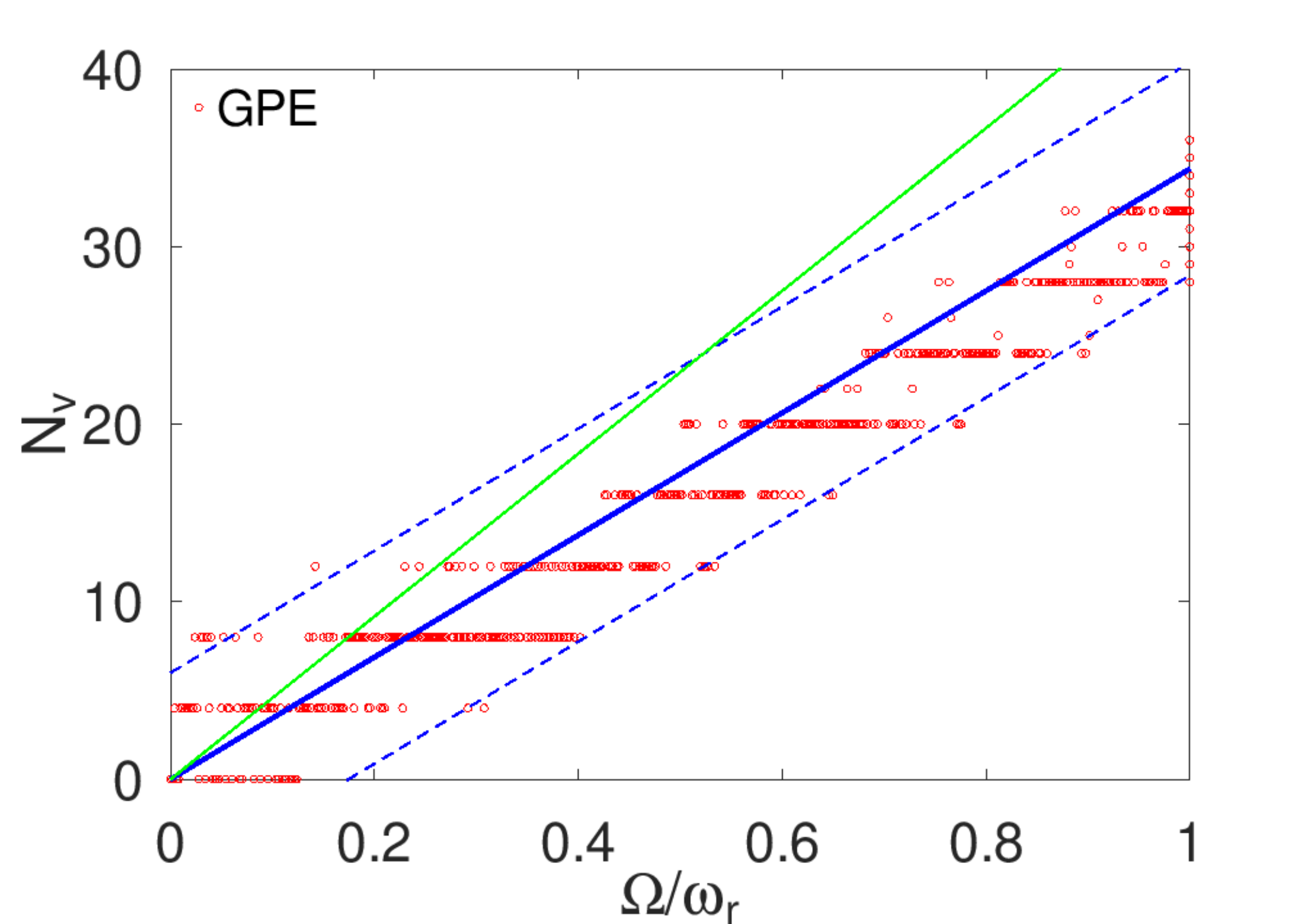}\\
 (b)\\
 \includegraphics[width=0.85\columnwidth]{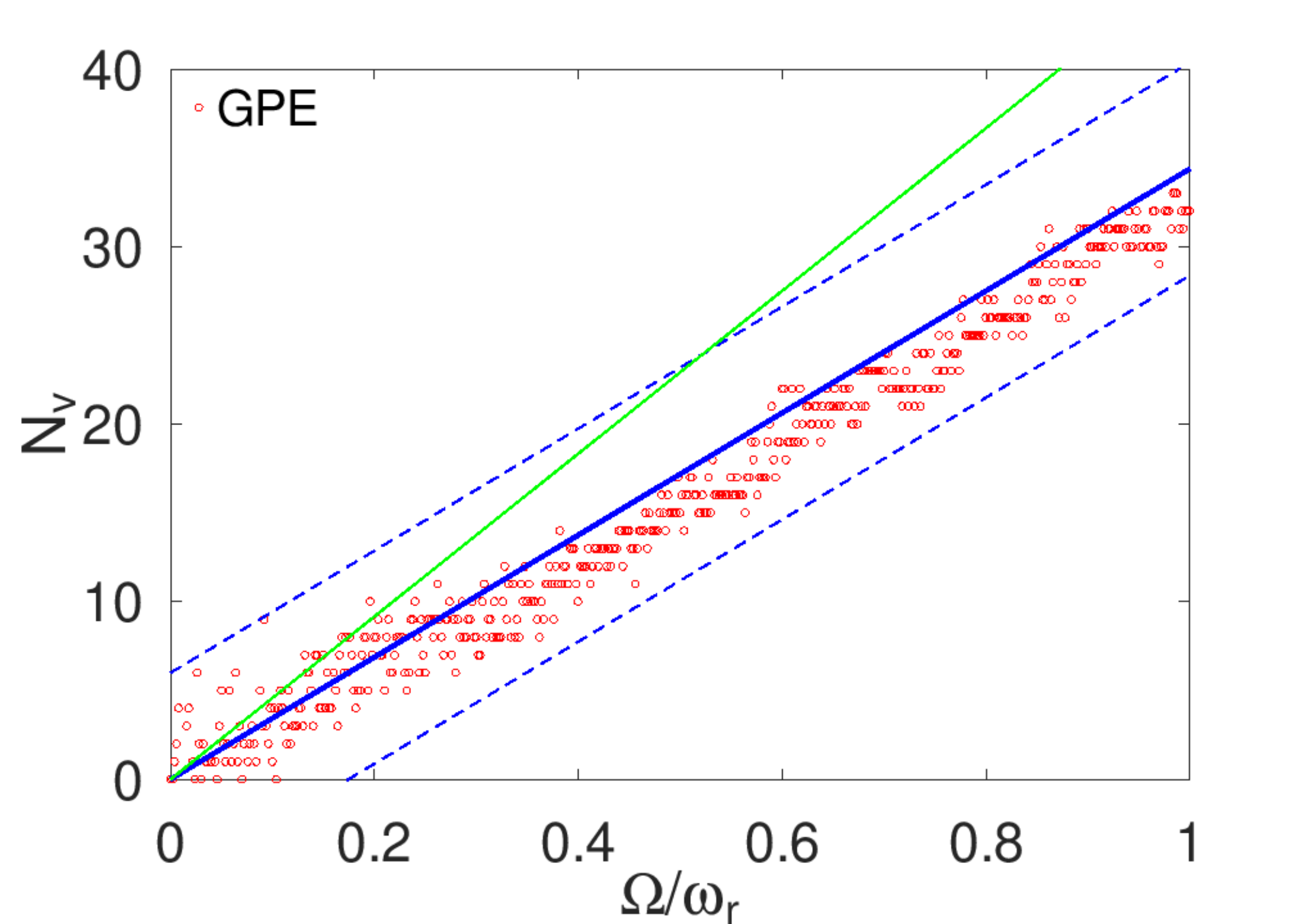}
\caption{Number of vortices $N_v$ as functions of the rotation frequency $\Omega$ for   $V_b=40\hbar\omega_r$.
The points  correspond  to time-dependent  GP results from   Eq. (\ref{GProtdin}).
The  top  panel (a) corresponds  to  simulations using small perturbed
 initial populations and the points in  the bottom panel  (b) were obtained  using a 
slightly shifted harmonic potential,    $\frac{ 1 }{2 } m \left[
    \omega_{r}^2 ( x-0.1 l_r)^2 +   \omega_{r}^2 ( y-0.1 l_r)^2 
    + \omega_{z}^2 z^2 \right] $. The lines are the same as in Fig. \ref{fig:Vb75}.
\label{fig:7}}
\end{figure}

The numerical simulations are performed  by solving  the 3D  time-dependent GP equation,
\begin{equation}
i \hbar  \frac{\partial }{\partial t}  \psi_{\text{GP}}(\mathbf{r},t) =
\left[ \hat{H}_0 +
g \, N|\psi_{\text{GP}}(\mathbf{r},t)|^2 - { \mathbf{\Omega(t)}}\cdot {\hat{ L}} \right] \psi_{\text{GP}}(\mathbf{r},t),
\label{GProtdin}
\end{equation} 
using as the  initial order parameter, $
\psi_{\text{GP}}(\mathbf{r},0)$,  the ground state of the
nonrotating system.  The  rotation angular  frequency  $
\mathbf{\Omega(t)}= \Omega(t)  \hat{ z}$  is  linearly increased
in time from zero up to  $ \omega_r  \hat{ z}$,   with a ramping
time interval $\tau= 10^3/\omega_r$.  We  change the lattice
intersite distance to  $d=3 l_r$ to simplify the notation, and
use a less confining harmonic    potential with
$\omega_r=2\pi\times 10$ Hz  and $\omega_z=2\pi\times 20$ Hz.
The number of sites is increased to  $N_s=16$ and the total
number of particles to $N= 4 \times 10^4$.  In Figs.
\ref{fig:Vb75} and \ref{fig:7} we show the number of vortices
nucleated within a lattice  as a function of $\Omega$, for $
V_b=75 \hbar\omega_r $ and $ V_b=40 \hbar\omega_r $,
respectively.  For both barriers the system is formed by WLCs,
being the barriers $ 1.3$  to $1.9$   larger than  the chemical
potential of the ground state. In order to observe how  the
results are  affected when the  four-fold  symmetry is slightly
broken, a fact  that can also occur  in an experiment, we discuss
two possibilities shown   in Fig. \ref{fig:7}.  In  the upper
panel we present the results when a small perturbation in the
initial populations of the ground state  has been applied, and in
the bottom one,  we perform a slight displacement $ \delta
\mathbf{r} = 0.1 l_r (\hat{x}+ \hat{y})$ of the axis of the
harmonic potential.  One can observe that a number of nucleated
vortices different from a multiple of four is now allowed, and
hence a smaller spreading of the points with respect to the line
that estimates the mean value is attained. In each case we also
draw the estimate $N_v^F = m \Omega d^2 N_s /   \pi \hbar $
which corresponds to $N_v^F = m \Omega R( \Omega=0)^2 / \hbar $
given in Ref.  \cite{williams10}.
We note that, when drawing the points of Figs.  \ref{fig:Vb75}
and \ref{fig:7}, we count the total number of GP  vortices by
adding the vortices and subtracting  the antivortices, both
obtained by the plaquette method, up to the borders of the
lattice. However, at such borders, some fluctuations appear which
give rise to the points that exceed the upper and lower bounds.

From our numerical calculation it may be seen that, for the
present dynamical  vortex nucleation process, the Feynman's rule
can be improved when applied to  an optical  lattice. Moreover,
although our approximation  has been derived for a different
system, it is easy to verify that it better describes  the
experimental data for the higher barrier of Ref.
\cite{williams10}, which correspond to WLCs.

In the next section we will see that the outer sites do not
increase their populations as expected from the centrifugal
distortion, and hence support  the fact of working with a fixed
number of sites during  the whole dynamics.  It is important to
distinguish our dynamical nucleation process from the theoretical
study \cite{ka11} where the dissipative parameter allows the
system to relax into a wider lattice and hence the number of
nucleated vortices increases  faster with $\Omega$.  Another
difference is the critical rotation frequency  $\Omega_c/
\omega_r \simeq 0.2 $ needed in Ref. \cite{ka11} to start
nucleating vortices.

In Figs. \ref{fig:Vb75} and \ref{fig:7} one can observe that the number of nucleated
vortices exhibits   fast variations as a function of the rotation
frequency. Such rapid variations are  related to the fact that a  self-trapping (ST) 
dynamics, with very short time periods, is triggered, which in turn
provokes a rapid movement of vortices.  In the following subsection we
will analyse such a vortex dynamics.

\subsection{Vortex dynamics}

In a homogeneous BEC, the vortex velocity is determined by the
background superfluid  velocity through  the Magnus force
\cite{don91,am80,sheehy04}.  Whereas in a nonhomogeneous BEC, the
vortex velocity acquires an additional term which depends on the
density gradient \cite{sheehy04,nil06,jezcat08,jez08} and on the
form of the vortex core \cite{jezcat08}. In systems where the
order parameter can be written in terms of LFs, an instantaneous
passage of a vortex can be easily predicted.  However, in these
cases, no relation between the velocity of the vortex and the
macroscopic coordinates, of the type of Eq. (\ref{vortimel}), has
been established.  In particular, in nonrotating double-well
systems, such passage of vortices across  the junctions has been
related to the presence of phase slips \cite{abad15} which occurs
in the  ST  regime \cite{smerzi97,ra99}. 
Their motion turn to be much faster than the variation of the
macroscopic coordinates which describe the relevant dynamics.  We
recall that the ST regime is characterized by a phase difference
time derivative that never vanishes, and is referred to  as
running phase \cite{smerzi97,ra99}.  It is worthwhile noting that
the ST behavior is also encountered in lattice systems
\cite{stop}, showing the same phase difference behavior.  In this
section, we will show that in rotating systems the GP vortex
dynamics is in close connection with the evolution of the phase
difference between neighboring sites, as stated in  Eq.
(\ref{vortimel}), and hence they share the same time periods.

\begin{figure}
 \centering
 \hspace*{2cm}(a)\\
 \includegraphics[width=0.75\columnwidth]{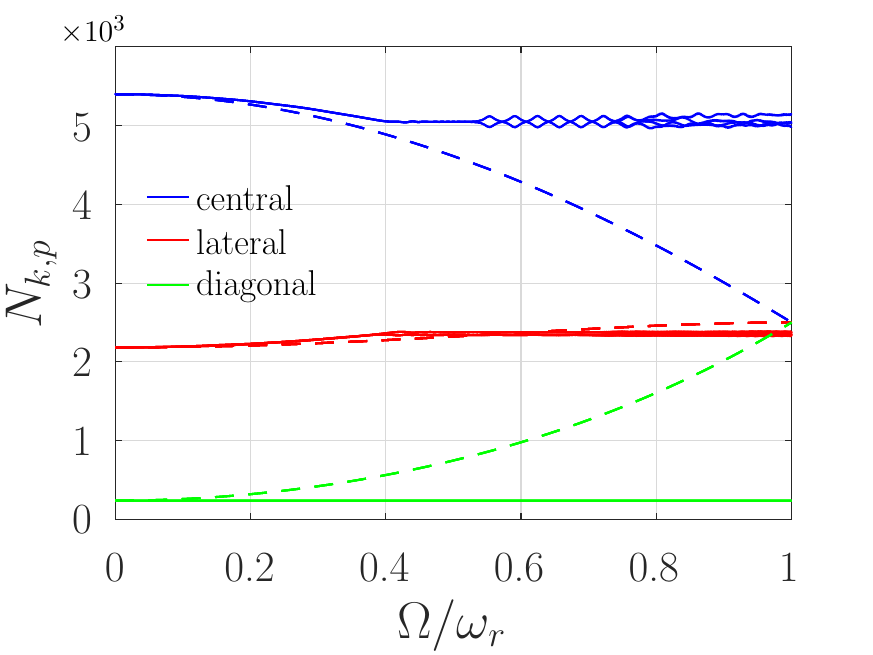}\\
 \hspace*{2cm}(b)\\
 \includegraphics[width=0.75\columnwidth]{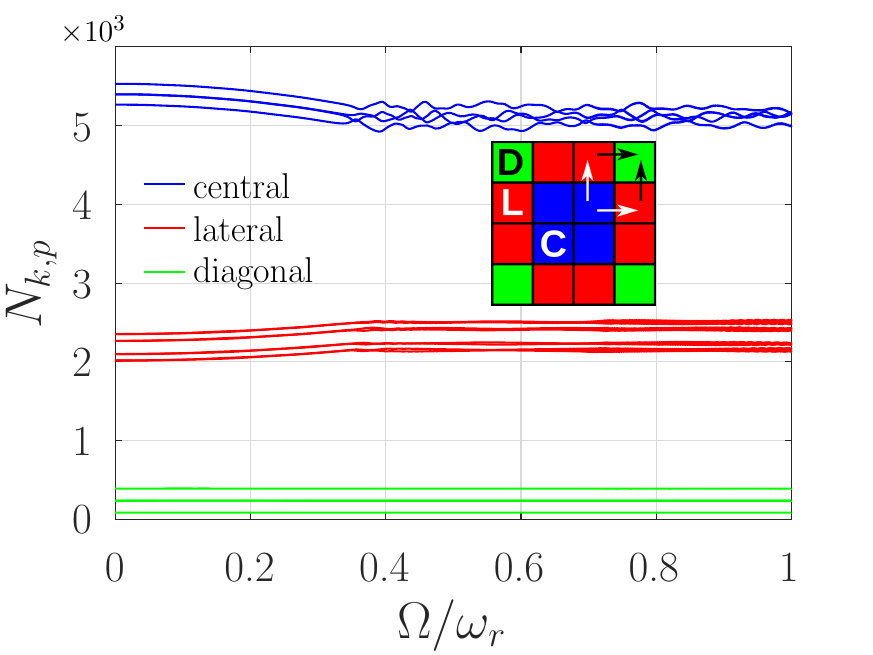}
 \caption{\label{fig:Ns40} Site populations  $N_{k,p}$  as functions of $ \Omega $, which  are depicted with
solid lines, were obtained by solving  Eq. (\ref{GProtdin})
during the time sweep of the frequency,  for
$V_b=40\hbar\omega_r$.   The top (a)  and bottom  (b)  panels
correspond to the same cases of Fig. \ref{fig:7}.  The blue
  (upper) lines, the red (middle) lines, and the green (lower)
  lines correspond  to the number of particles in the  central,
  lateral, and diagonal  sites, respectively.  The correspondence
  between the  indices $k,p$ and the different sets of sites is
  explained in the text.  The dashed lines in the top panel
  depict the GP stationary populations at the corresponding
  static  frequency, obtained via a minimization procedure.}
\end{figure}

In stationary conditions, attained from a  GP  energy
minimization procedure, each phase difference  and  population
should  remain  static. However, the static  population in each
site should differ for distinct constant  frequencies  due to the
centrifugal force.  Here, we may distinguish three sets of sites
according to their center-of-mass distance to the rotation axis,
namely, the central (C), lateral (L), and diagonal (D) ones,
consisting of four, eight and four sites, respectively (see inset
of Fig. \ref{fig:Ns40}). In the notation of  Eq. (\ref{orderp}),
the central sites  labels $k$ and $p$ are given  by the
combinations of the numbers $ \{2;3\}$, the diagonal by $
\{1;4\}$, and the lateral by  the rest ones.  Hence, the
centrifugal force should push particles from central to lateral
sites, and from lateral to diagonal ones as marked with arrows in
the right corner of inset of Fig. \ref{fig:Ns40}.  In the top
panel of Fig. \ref{fig:Ns40}, we show as dashed lines the number
of particles in the different types of sites in stationary
conditions as function of the static rotation frequency.  It may
be seen, for example, that all sites acquire the same number of
particles when $\Omega= \omega_r$, due to the full compensation
of the harmonic trapping with the centrifugal force.

\begin{figure}[h!]
\centering
  \hspace*{-2.5em}\includegraphics[width=\columnwidth,clip=true]{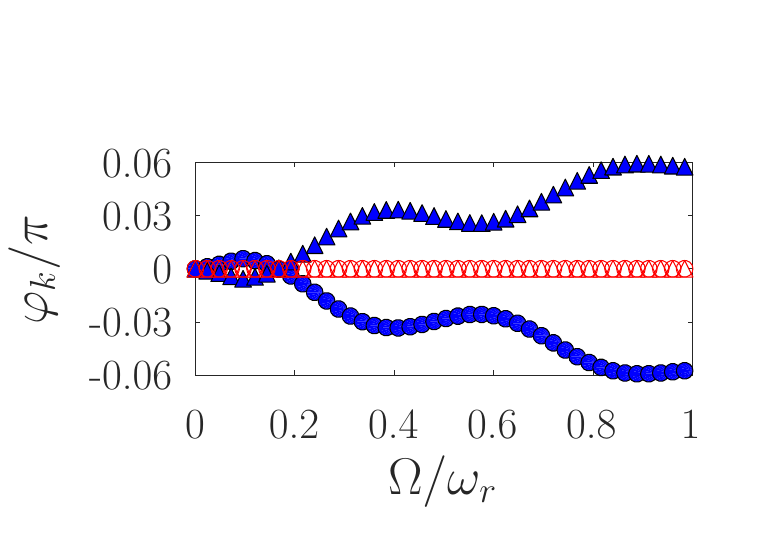}\\[-5pt]
\includegraphics[width=0.95\columnwidth,clip=true]{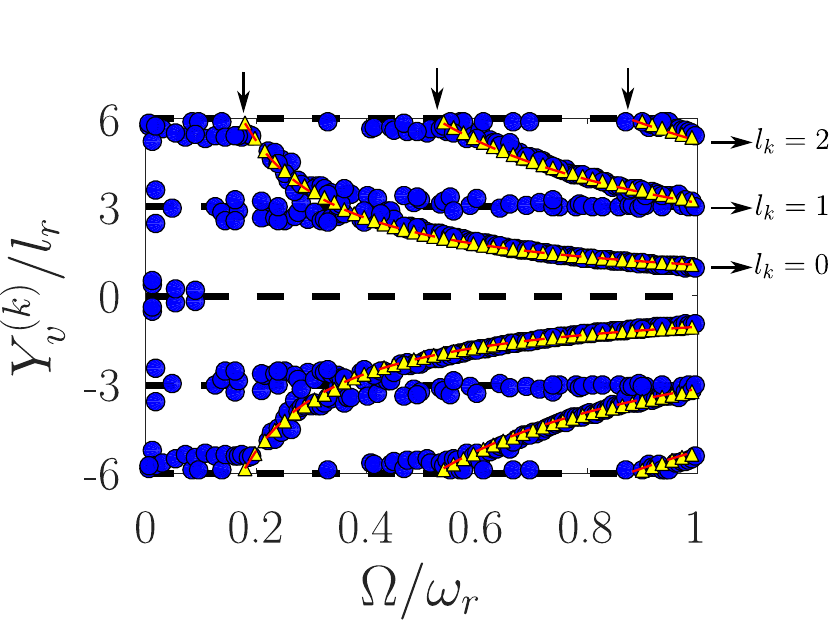}\\[-3pt]
\includegraphics[width=0.95\columnwidth,clip=true]{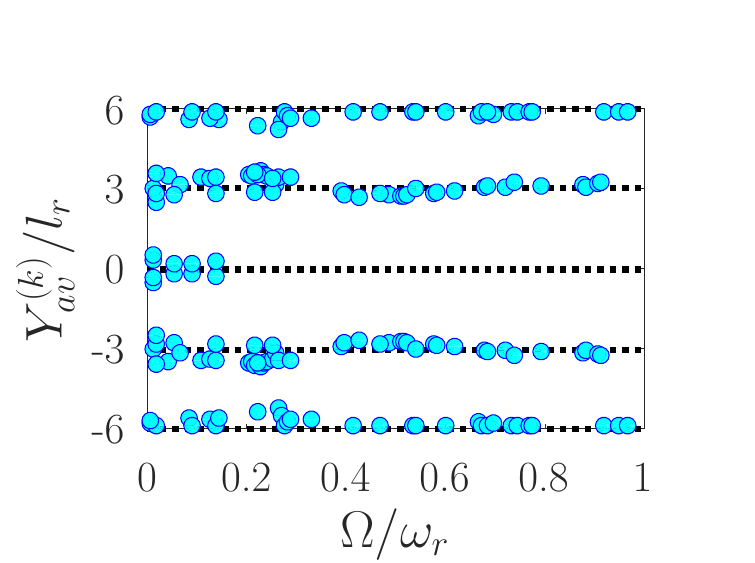}
\caption{\label{fig:vortex_at_0} Top panel:  GP phase differences
between sites at both sides   along $x=0$,  $\varphi_{k,p} =
  \phi_{k-1,p} - \phi_{k,p} $, which are given for $k=3$ and
  $p=1$ to $4$,  as functions of $\Omega$, for $V_b=75
  \hbar\omega_r$.  The  phase difference between the lateral
  sites for $y>0$ ($y<0$) with $p=1$ ($p=4$)  is depicted with
  blue filled triangles (circles), whereas the difference between
  the central sites for $y>0$ ($y<0$) with $p=2$ ($p=3$)  is
  depicted with red hollow triangles (circles).  Middle panel:
  Vortex positions along the $x=0$ low density path. The
  positions extracted from the order parameter  of  Eq.
  (\ref{GProtdin}), obtained by the plaquette method, are
  depicted  with blue dots, whereas the red dashed line and the
  yellow triangles, almost superimposed on the blue dots,
  correspond to Eq. (\ref{vortimel})  replacing  $\varphi_k(t)$
  by zero and by  the GP  $\varphi_{3,p}(t)$  of the top panel,
respectively. Bottom panel: Antivortices obtained through
GP-simulations.  } \end{figure}

For the barrier of Fig. \ref{fig:Vb75}, we found that the number of
particles in each site remains fixed at their initial value during all
the time evolution. Whereas, for the lower barrier, a
different evolution of the populations is encountered. In
Fig. \ref{fig:Ns40} we show  the  low barrier height case  with 
$V_b=40 \hbar\omega_r $, which is nearer to the chemical potential
which is about $\mu \simeq 30 \hbar\omega_r $.  It may be seen that
there exists an initial movement of particles to the more external
sites up to $\Omega \simeq 0.4 \omega_r$.  In particular, in that
range of  frequencies, a sizable number of particles
move from the central sites to the lateral ones. For larger values of
$\Omega$, the populations only  perform  small  oscillations around some fixed
numbers.

In what follows we will see how the populations $N_{k,p}$, the
phase differences, and  the vortices  evolutions are related
when the  time-linear  ramp of the rotation frequency  is
applied.  As a consequence of the  phase differences evolution a
dynamics of vortices   appears as expected from Eq.
(\ref{vortimel}), which  exhibits distinct behaviors.  More
precisely,  for neighboring sites that belong to the same set
(C, L or D), very small variations of the  phase differences are
observed, and hence, the vortices located   between such sites
move slowly.  Whereas, when the neighboring sites belong to
different sets, and the populations turn to be far from the
stationary configuration values,   running phase differences,
with related  fast vortex dynamics,  develop.  Hence, the fast
variation of the phase differences, conected to a ST regime,
seems to be the responsible for the permanence  of almost all
the particles within the same sites of the nonrotating system.
Such an observation upon the atoms remaining within the initial
confinement has also been reported in the experimental work
\cite{williams10}.

We first study the dynamical nucleation process  along the low density
paths that separate sites belonging to the same set, as, for example,
the $x=0$ path, where the phase differences are calculated between two
central sites and two lateral ones.
In the notation of  Eq. (\ref{orderp}) such  phase differences are given by 
$\varphi_{3,p}(t)=\phi_{2,p}(t)- \phi_{3,p}(t)$ with $p=1$ to $4$.
From  the top 
panel of Fig. \ref{fig:vortex_at_0}  it may be seen that  the phase
differences remain almost vanishing, since 
$ |\varphi_{k}(t)|/\pi \le 0.06$,  and thus a reasonable estimate for the
position of vortices can be obtained using Eq.  (\ref{vortimel})  with
$ \varphi_k=0$, for all $p$ values. 
In the middle  panel of
Fig. \ref{fig:vortex_at_0} we display the vortex positions.
Focussing on the vortices with  $Y^{(k)}_{v}>0$, with $k=3$,  we
may see that the positions    obtained  from  Eq.
(\ref{vortimel})  with  $ \varphi_k$ replaced by the GP results
$ \varphi_{k,p}(t) $  (for  $k=3$ and  $p=1,2$ values) and with $
\varphi_k=0$,  where we have used $l_k=0,1$, and $2$, are
superposed, and hence the variation of the phase differences has
no sizable effect on the vortex nucleation at the axes.
Moreover,  both curves  are in good accordance with the points
extracted  from the  time-dependent GP simulations using the
plaquette method.  We note that at the end of the evolution there
are three  vortices  located at the points  marked with
horizontal arrows in the middle panel,    that have  entered the
lattice  at the frequencies  indicated with vertical arrows.
Given that  the largest  $l_k  $ value is   $l^{(L)}_k  =2$  the
number of vortices that entered through  the top entrance is in
fact  $N^{(k)}_v=l^{(L)}_k +1=3$,  as stated above   Eq.
(\ref{niv}).  The reflection symmetry observed  in such a panel
is a direct consequence of the four-fold symmetry of the lattice.
Then, by  the four entrances located at the axes, during the
whole evolution  $12$ vortices  penetrate in the system.

It is important to mention that many vortex-antivortex pairs are
created and annihilated instantaneously throughout the GP dynamics,
specially near the intersections of the low density paths. Hence, as
we are depicting only the vortices, in the middle  panel, some points
that appear occasionally in the graph correspond to such
vortex-antivortex fluctuations, and they do not contribute to the net
number of nucleated vortices, as they cancel each other. In the bottom 
panel of Fig. \ref{fig:vortex_at_0} we show the antivortices
positions. Such antivortices   appear,
as we have stated  around  the intersections of the paths,  
where the
approximation of Eq. (\ref{vortimel}) is not appropriate  for predicting  the existence
of vortices, since the  four LFs should be taken into account. 
We want to mention that by analysing the
three-dimensional structure of such vortex-antivortex pairs, we have
observed that they do not correspond to straight vortex lines and in
some cases they form part of a vortex ring.

\begin{figure}[h!]
	\centering
	\includegraphics[width=\columnwidth,clip=true]{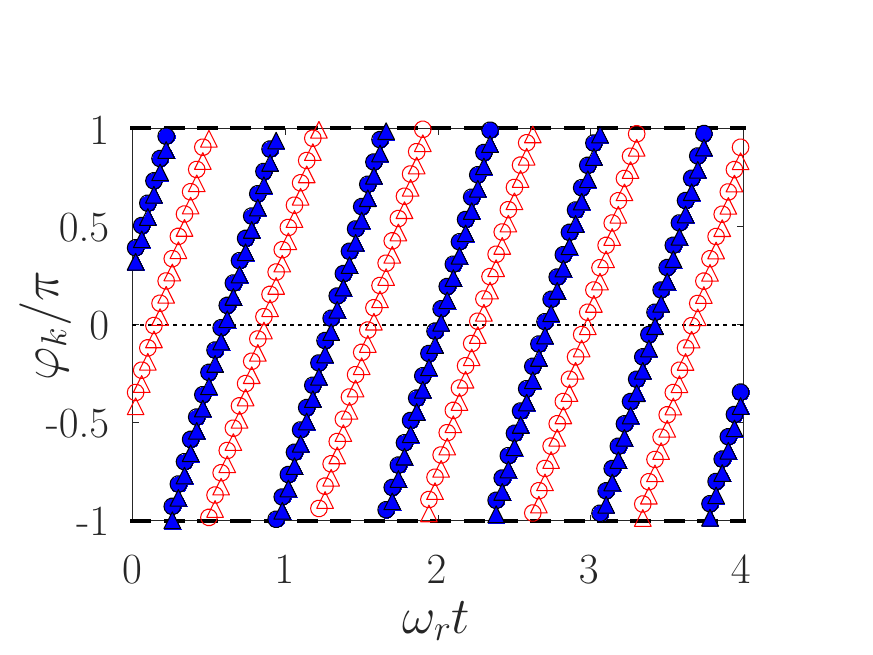}\\
	\includegraphics[width=\columnwidth,clip=true]{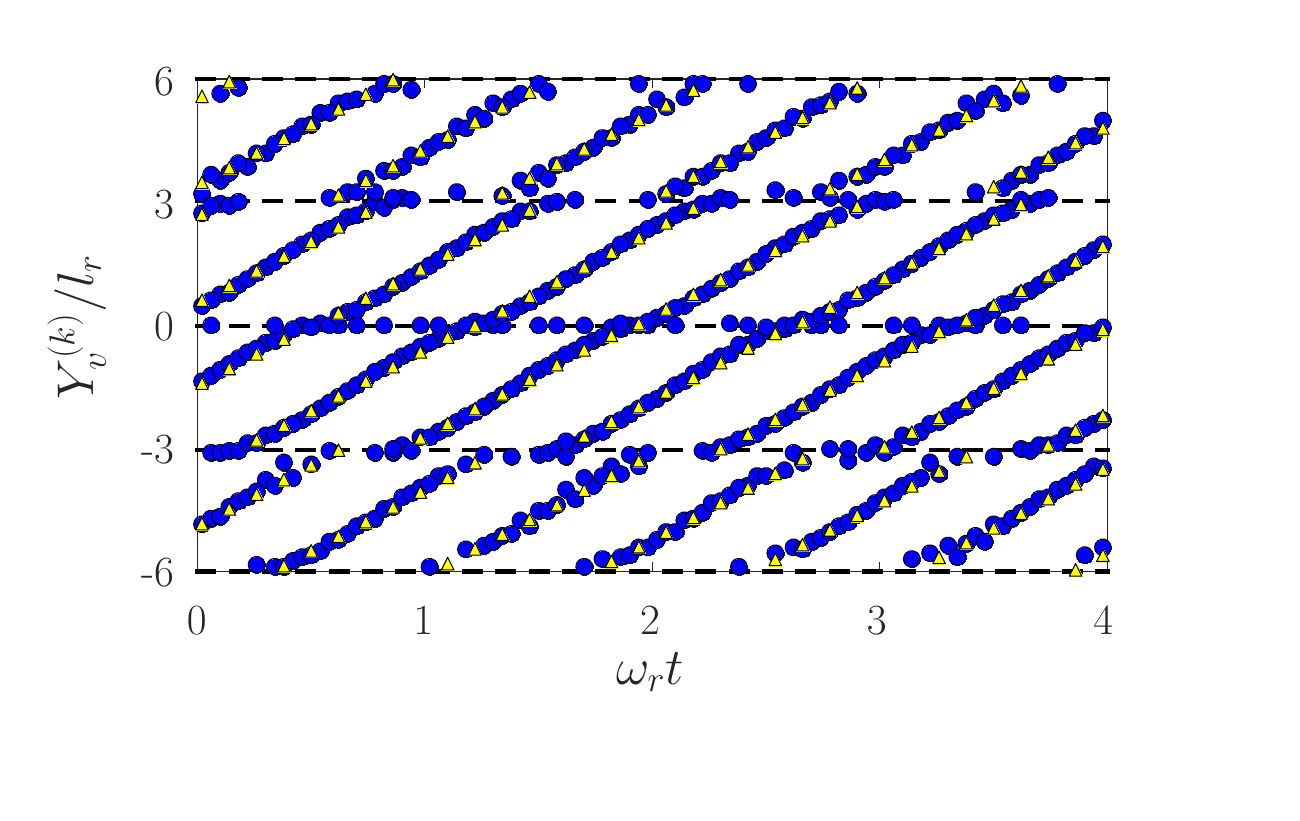}
	\caption{\label{fig:vortex_at_3}
	Top panel:  The phase
	differences as function of time  between both sides  of the  line  $x/l_r=-3$,   $\varphi_{k,p}(t)=\phi_{k-1,p}(t)- \phi_{k,p}(t)$, 
	for $k=2$ and $p=1-4$. The phase difference 
	between the diagonal and lateral
	sites $\varphi_{2,1} $ ( $\varphi_{2,3} $),  for $y>0$ ($y<0$)  is depicted with blue filled triangles
	(circles), whereas the difference between the lateral and central
	sites for $y>0$ ($y<0$) with index $p=2$ ($p=3$)  is depicted with red hollow triangles
	(circles).  Given the high symmetry of the system both curves with  blue (red) points  are superimposed.
	Bottom panel:  The vortex positions $Y_v^{(k)}$ as function of time are depicted with  blue dots  for 
	time-dependent  GP simulations  by solving Eq. (\ref{GProtdin}), 
	and the yellow triangles  (almost on top of the blue ones)  are obtained through
	Eq. (\ref{vortimel}) replacing  $ \varphi_{2}(t)$ by  the GP results  $ \varphi_{2,p}(t)$, with $p=1$ to $p=4$,
	which yield the different  $Y_v^{(2)}$ values  from top to bottom. 
	The  barrier height  is $V_b=75 \hbar\omega_r$ and the
	time interval has been extracted after the linear ramp was switched
	off. }
\end{figure}

For the paths that separate   neighboring sites belonging to
different sets (C, L, or D), the time-dependent  nucleation of
vortices is not only due  to the process observed at the axes.
The phase differences time  variations, between the centers of
such sites, can also provoke  the penetration and departure of
vortices,  and hence change the values  of  $l^{(L)}_k$.  In
particular,   when the  on-site populations depart sizably from
their stationary values, the vortex  nucleation  process is
combined with  a rapid motion of vortices.  The fast vortex
dynamics  seems to be related   to the fact that the macroscopic
coordinates  enter  a self-trapping regime characterized by an
almost constant occupation number and a running phase difference
between such neighboring sites.  From Eq. (\ref{vortimel}) one
can infer that an increasing (decreasing) running phase should
produce a monotonous increasing (decreasing) vortex coordinate.
In our system, such dynamics develops along the four  low density
paths defined by  the straight  lines  $ x= \pm 3 l_r$ and  $ y
=  \pm 3 l_r $, which separate sites of distinct sets. In Fig.
\ref{fig:vortex_at_3}, we illustrate such behavior along the path
located at $x =-3 l_r$, when the linear  ramp has been switched
off, and hence around six vortices have been  nucleated.  In the
top panel we show the phase differences
$\varphi_{k,p}(t)=\phi_{k-1,p}(t)- \phi_{k,p}(t)$ for $k=2$ and
different $1 \leq p \leq 4$ values,  from time-dependent  GP
simulations.
In the bottom panel we show the  vortex coordinate extracted from
time-dependent  GP simulations, through the plaquette method.
Almost on top of such points, indicated with yellow triangles, it
can be viewed    the points estimated by Eq.  (\ref{vortimel})
using the time-dependent  GP-phase differences, showing the  very
good agreement between both approaches.  Hence, one can conclude
that such vortex dynamics is ruled by the phase differences
evolution, rather than by density gradients which become useful
in other type of systems.  Furthermore, such vortex dynamics is
much more rapid than the typical timescales involved in the whole
process  provided by the linear ramp.  In the bottom panel it may
be seen that vortices enter the lattice from the bottom entrance
and depart from the top one at different times,  which explains
the rapid variation of the nucleated vortices inside the lattice.
It is  worthwhile recalling  that if the  four-fold symmetry  is
broken, as in the case of the  bottom panel of Fig. \ref{fig:7},
the fluctuation on the total number of vortices has shown to
exhibit  a smaller amplitude. A plausible  explanation of such an
effect could be that  the vortices  that move along the four
straight lines $ x= \pm 3 l_r$ and  $ y  =  \pm 3 l_r $,  do not
enter, or exit, the system   simultaneously.

\begin{figure}[h!]
	\centering
	\includegraphics[width=0.85\columnwidth,clip=true]{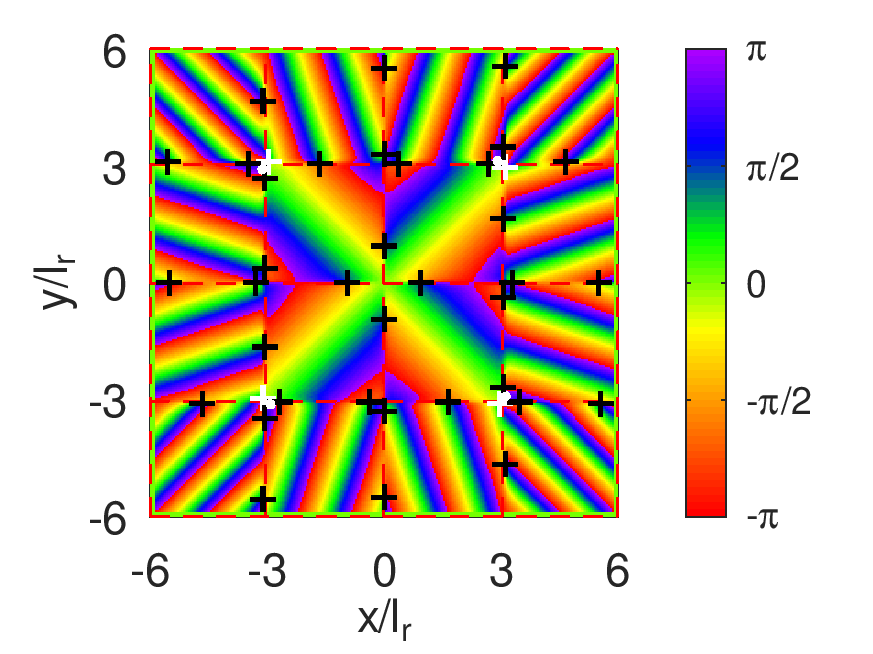}\\
  \includegraphics[width=0.85\columnwidth,clip=true]{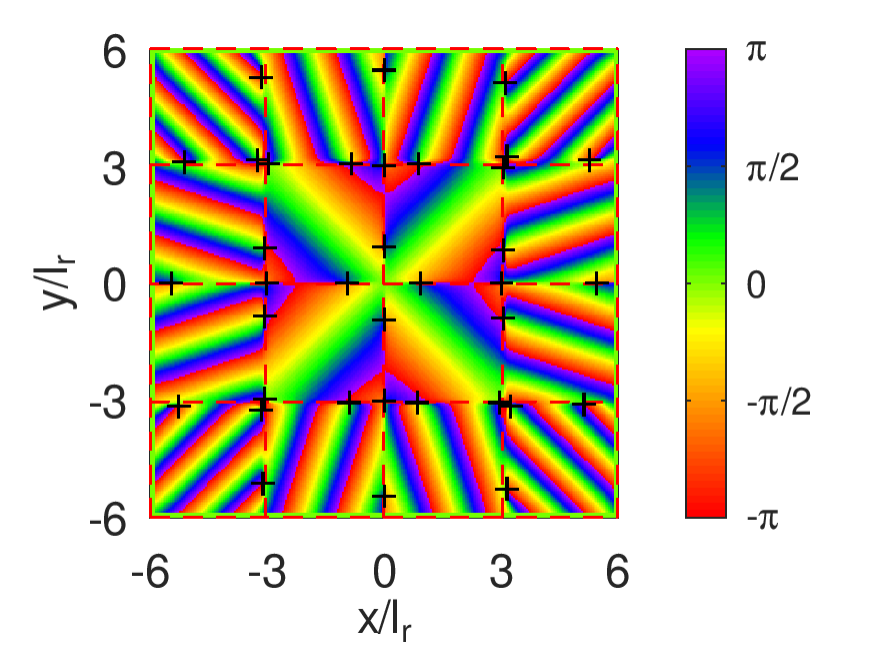}\\
  \includegraphics[width=0.85\columnwidth,clip=true]{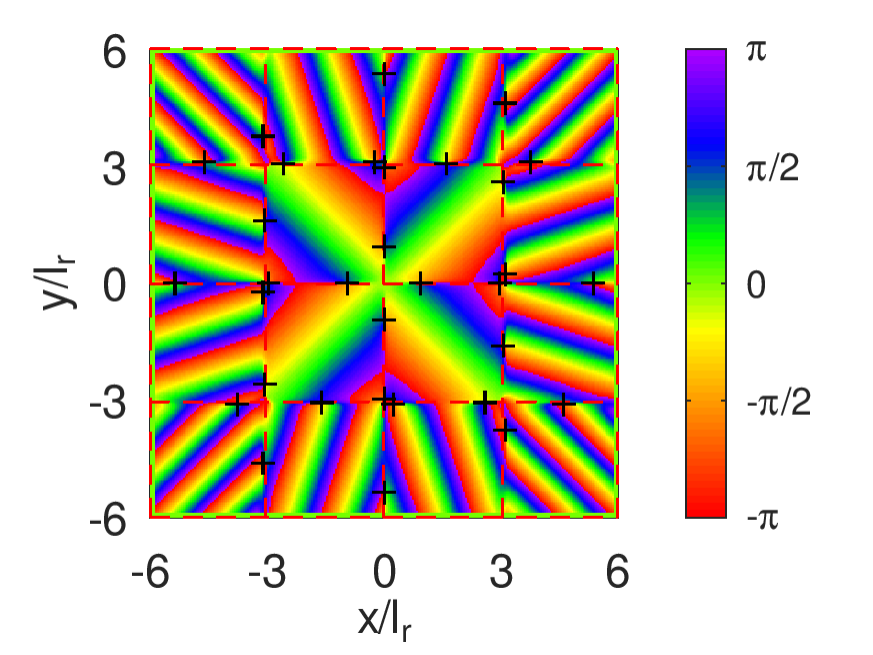}
  \caption{\label{fig:vortex_at_xy} The phase of the  GP order
  parameter, obtained by solving  Eq. (\ref{GProtdin}),  is
  depicted in colors at the $z=0$ plane.  The corresponding
  vortex locations, obtained by the plaquette method, are marked
  with plus signs, whereas the antivortices with circles.  The
  nucleated vortices are marked with black symbols, whereas
  vortex-antivortex pairs, in the top panel around $|x|=|y|= 3
  l_r$,  are depicted with white symbols. The barrier height
  corresponds to  $V_b=75 \hbar\omega_r$.  The three times were
  selected at the end of the ramp, at $\omega_r t=999.8$ (top
  panel), 1000.0 (middle panel), 1000.2 (bottom panel).  In each
  panel the  color scale  corresponds to $\arg{(
  \psi_{\text{GP}}(\mathbf{r},t)) }$ at the selected time.}
\end{figure}

Finally, to better visualize the vortex dynamics  in Fig. \ref{fig:vortex_at_xy} we 
 mark with plus signs the positions of vortices obtained by time-dependent GP simulations,  in
the $ z=0 $ plane, at three times around
$t \simeq 1000 \omega_r^{-1} $. In such figure, we also  show  in colors, the phase of the GP
order parameter where  one can first view
that the phases are, in fact, linear on such coordinates at each site,
as we have assumed, and that the phase gradient increases with an
increasing center-of-mass distance to the rotation axis. In other
words, the central, lateral, and diagonal sites have increasing phase
gradients in such an order. Secondly, the vortices are clearly located
along the six paths defined by the straight lines, 
$ x/l_r= -3,0, 3$  and $ y/l_r = -3,0,3$. Along the axes $ x=0$
and $y=0$ the vortices only perform very  small oscillations, and hence
their total number remains fixed at six vortices per axis in the time
interval considered, and the  positions are in agreement with 
  those  nucleated   at the end of 
the evolution of Fig. \ref{fig:vortex_at_0}, which are marked 
with horizontal arrows at the middle panel of such a figure. 
 On the other hand,
for the other four  straight lines  $ x/l_r= \pm 3$ or  $ y/l_r= \pm 3$, the running  phase difference, 
  shown in Fig. \ref{fig:vortex_at_3},   generates a
vortex dynamics, which can involve either the  ingress or   the exit of  vortices from the lattice.
For example,   at $ x =-3 l_r $ it may be
seen, by following the movement of the vortex which is
around $y=5l_r$ at the top  panel of  Fig. \ref{fig:vortex_at_xy}, that it moves in the positive 
$y$-direction to its position in the middle panel, and finally it
disappears at the bottom panel. Hence, one has six vortices moving in
such direction, except in the lower panel where only five remain. Once
more, such dynamics explains the change, from 36 to 32,  of the total number of
vortices nucleated, in a so short time interval. Using our estimate  of Eq. (\ref{nva}), 
the number of vortices  yields 34.4 in
contrast to Feynman's rule applied to our 16 sites-system, which reads 45.8.
 The top  panel also
illustrates the appearance of vortex-antivortex pairs, 
around the points that satisfy $ |x|=|y|=3 l_r$, that disappear
in the middle panel. We note that  such fluctuations are concentrated in the
lower density regions.

\section{Conclusions}

For a four site system which forms a ring lattice, by using an
accurate expression for the phases of the on-site localized
functions,  we  have analytically obtained  the arrays of
vortices for the four different stationary states with  a given
rotation frequency.  By comparing with the  GP results, we  have
shown that our analytical estimates  for such  vortex positions
turn to be very accurate.

Hence,  in a second step,  making use of the same type of
expressions,  we have investigated  a vortex nucleation process
for a square lattice  with a  larger number of sites.  Such a
process consists on applying a  a time-linear ramp of the
rotation frequency, similar to the method used in the experiment
of Ref.  \cite{williams10}. 
By  analysing such a  dynamical process, we found that the
on-site populations perform   small oscillations around   numbers
very close to  the initial ones, and that the  phase differences
between different types of sites,  exhibit a running phase
behavior as  function of time,  typical of a ST  regime. As a
consequence of such a behavior, the expanding  effect of the
centrifugal force  was  cancelled. An analogous observation, for
WLCs,  has been reported in Ref.  \cite{williams10},  where the
experimental  results   do  not reflect an increasing
Thomas-Fermi confinement radius as expected from a centrifugal
distortion.

With respect to the number of nucleated vortices  we have
obtained a linear dependence on the rotation frequency similar to
the estimate $N_v^F = m \Omega d^2 N_s /   \pi \hbar $  used in
Ref. \cite{williams10} for WLCs.  However, our prediction differs
from such an estimate, given  that,  instead of being
proportional to the number of involved sites,   it turns to be
proportional to $ N_s - \sqrt{N_s}$. By introducing the
parameters of the experiment of  Ref. \cite{williams10}  in our
formula, it  seems to better reproduce their experimental
results.  Although, our approach was derived for a specific
system, we believe that the new  factor   which is related to the
number of entrances,  should also be present   in the  vortex
nucleation estimates  on other types of lattice potentials.

Finally, the nucleation of vortices in other systems of current
interest such as a rotating supersolid dipolar gas
\cite{roccu20,gal20} with regular patterns in the density could be
suitable for applying the present approach.  Given that each droplet
exhibits an almost axial symmetry around an axis parallel to the
rotation axis, when subject to rotation its velocity field can be
approximated by an homogenous one \cite{rot20}.  Then, the rotation induced
phase on the droplet should have a linear dependence on the
coordinates, which constitutes the main requirement for the
application of the present treatment. Since rotation may compromise
the density pattern of the nonrotating ground state
\cite{gal20,gal22}, the study should contemplate the possible changes,
when varying the rotation frequency, in the number of droplets and
their distribution.

\begin{acknowledgments}

D.M.J. and  P.C  acknowledge CONICET for financial
support under PIP Grant No. 11220150100442CO. 
	P. C. acknowledges for  financial support from the Universidad de Buenos Aires (grant UBACyT 
	20020190100214BA) and the Consejo Nacional de Investigaciones Científicas y Técnicas (grant PIP  11220210100821CO).
\end{acknowledgments}

\end{document}